\documentclass[apj,letterpaper]{emulateapj}

\usepackage{rotating}

\gdef\Ja{$J_1$}
\gdef\Jb{$J_2$}
\gdef\Jc{$J_3$}
\gdef\Ha{$H_1$}
\gdef\Hb{$H_2$}

\def\sigM{\sigma_{\log M}}

\def\Mmin{M_{\rm min}}
\def\M1{M_1^\prime}
\def\Msun{M_\odot}
\def\Mcut{M_{\rm cut}}
\def\hMsun{$h^{-1}M_\odot$}
\def\erf{{\rm erf}}

\def\xir{$\xi(r)$}
\def\wt{$w(\theta)$}

\slugcomment{Accepted for publication in the
Astrophysical Journal}

\shorttitle{Galaxy clustering in the NEWFIRM Medium Band Survey}
\shortauthors{Wake et al.}

\begin{document}

\title{Galaxy clustering in the NEWFIRM Medium Band Survey: the relationship between stellar mass and dark matter halo mass at $1 < z < 2$}

\author{David A.\ Wake\altaffilmark{1},
Katherine E.\ Whitaker\altaffilmark{1,2},
Ivo Labb\'e\altaffilmark{5,2},
Pieter G.\ van Dokkum\altaffilmark{1,2},
Marijn Franx\altaffilmark{3},
Ryan Quadri\altaffilmark{5,3,2,9},
Gabriel Brammer\altaffilmark{1,2},
Mariska Kriek\altaffilmark{7,4,2},
Britt F.\ Lundgren\altaffilmark{1},
Danilo Marchesini\altaffilmark{6,2},
and Adam Muzzin\altaffilmark{1}
}

\altaffiltext{1}{Department of Astronomy, Yale University, New Haven, CT 06520-8101}
\altaffiltext{2}{Visiting Astronomer, Kitt Peak National Observatory,
  National Optical Astronomy Observatory, which is operated by the
  Association of Universities for Research in Astronomy (AURA) under
  cooperative agreement with the National Science Foundation.}
\altaffiltext{3}{Sterrewacht Leiden, Leiden University, NL-2300 RA Leiden,
The Netherlands.}
\altaffiltext{4}{Department of Astrophysical Sciences, Princeton
  University, Princeton, NJ 08544.}
\altaffiltext{5}{Carnegie Observatories, Pasadena, CA 91101.}
\altaffiltext{6}{Department of Physics and Astronomy,
Tufts University, Medford, MA 02155.}
\altaffiltext{7}{Harvard-Smithsonian Center for Astrophysics, 60 Garden
Street, Cambridge, MA 02138.}
\altaffiltext{8}{European Southern Observatory, Alonso de C´ordova 3107,
Casilla 19001, Vitacura, Santiago, Chile.}
\altaffiltext{9}{Hubble Fellow.}

\begin{abstract}
We present an analysis of the clustering of galaxies as a function of their stellar mass at $1 < z < 2$ using data from the NEWFIRM Medium Band Survey (NMBS). The precise photometric redshifts and stellar masses that the NMBS produces allows us to define a series of stellar mass limited samples of galaxies more massive than $7\times10^9\Msun$, $1\times10^{10}\Msun$ and $3\times10^{10}\Msun$ in three redshift intervals centered on $z=1.1, 1.5$ and $1.9$ respectively. In each redshift interval we show that there exists a strong dependence of clustering strength on the stellar mass limit of the sample, with more massive galaxies showing a higher clustering amplitude on all scales.
We further interpret our clustering measurements in the $\Lambda CDM$ cosmological context using the halo model of galaxy clustering. We show that the typical halo mass of both central and satellite galaxies increases with stellar mass, whereas the satellite fraction decreases with stellar mass, qualitatively the same as is seen at $z < 1$. We see little evidence of any redshift dependence in the relationship between stellar mass and halo mass over our narrow redshift range. However, when we compare our measurements with similar ones at $z\simeq0$, we see clear evidence for a change in this relation. If we assume a universal baryon fraction, the ratio of stellar mass to halo mass reveals the fraction of baryons that have been converted to stars. We see that the peak in this star formation efficiency for central galaxies shifts to higher halo masses at higher redshift, moving from $\simeq7\times10^{11}$\hMsun~at $z\simeq0$ to $\simeq3\times10^{12}$\hMsun~at $z\simeq1.5$, revealing evidence of `halo downsizing'. Finally we show that for highly biased galaxy populations at $z > 1$ there may be a discrepancy between the space density and clustering predicted by the halo model and the measured clustering and space density. This could imply that there is a problem with one or more ingredient of the halo model at these redshifts, for instance the halo bias relation may not yet be precisely calibrated at high halo masses or galaxies may not be distributed within halos following an NFW profile. 

\end{abstract}


\section{Introduction}

Understanding the formation and evolution of galaxies in a cosmological context remains one of the most challenging problems in modern astrophysics. In the current cosmological framework, where the mass in the universe is dominated by cold dark matter (DM), luminous galaxies form at the centers of dark matter halos via the cooling and condensation of baryons \citep{White78,Fall80,Blumenthal84}. This means that the properties of galaxies are directly coupled to those of the dark matter halos in which they live.

If we wish to understand galaxy formation within this context, it becomes important to try to link the observed properties of galaxies, such as stellar mass or color, to the mass of the halos hosting galaxies with those observed properties, to better understand the physical processes involved. Making such a direct link can be achieved relatively easily in massive clusters of galaxies, with X-ray, Sunyaev-Zel'dovich effect and strong and weak lensing measurements, but it is much more challenging for less massive halos. Dynamical measurements of bound satellites \citep[e.g.][]{More10}, or strong \citep[e.g.][]{Auger10} and weak \citep[e.g.][]{Mandelbaum06} gravitational lensing, while effective techniques, are observationally expensive and have thus mainly been used for galaxies in the local universe, with only a few studies up to $z \sim 1$ \citep[e.g.][]{Heymans06,Conroy07}.

Measuring the spatial clustering of galaxies provides an alternative approach to relating galaxy properties to those of the DM distribution. More clustered populations must occupy regions of higher dark matter density ( i.e. more massive dark matter halos), than less clustered populations.
The desire to determine the link between galaxies and dark matter halos from clustering measurements has led to the development of the Halo Occupation Distribution (HOD) framework \citep{Jing98,Ma00,Peacock00,Seljak00,Scoccimarro01,Berlind02,Cooray02}. The HOD characterizes the statistical relationship between galaxies and dark matter halos by describing the probability that a halo of a given mass hosts a certain number of galaxies with a given property.  

The recent completion of large redshift surveys in the local universe such as the Sloan Digital Sky Survey \citep[SDSS;][]{York00} and the Two Degree Field Galaxy Redshift survey \citep{Colless01} have allowed precise measurements of the clustering of galaxies as a function of their intrinsic properties, such as luminosity, color, star formation rate, and morphology \citep{Norberg01,Norberg02,Zehavi02,Budavari03,Madgwick03,Zehavi05,Li06,Swanson08,Ross09,Loh10,Ross10,Zehavi10}. This has led to an established observational picture with galaxies becoming more clustered on all scales as their luminosity or stellar mass increases. As the color becomes redder, or the star formation rate decreases, the clustering strength again increases where the magnitude of the increase becomes larger on small scales. 

These relationships between galaxy properties and clustering strength can straightforwardly be interpreted in the framework of the HOD \citep[or closely related Conditional Luminosity Function, CLF;][]{Yang03}. Such analyses reveal an increase in the typical mass of the host halos as the galaxy stellar mass increases, and that the distribution of satellite galaxies in massive halos is a strong function of their color \citep[or star formation rate;][]{Yan03,Yang05,Zehavi05,Zheng07,Ross09,Zehavi10}. Such constraints on the relationship between galaxy properties and those of dark matter halos provide both insight into the physics of galaxy formation and particularly strong tests of any cosmological galaxy formation model.

Observations of galaxy clustering up to a redshift of one, from both large spectroscopic and photometric redshift surveys, appear to show similar trends as those observed in the local universe. More massive/luminous galaxies show stronger clustering and are thus associated with more massive halos, and the relationships between color and clustering seems to persist \citep{Coil04,LeFevre05,Phleps06,Coil06,Pollo06,Coil08,Meneux08,McCracken08,Meneux09,Simon09,Abbas10}. Again the HOD has been effectively used to interpret these measurements \citep{Yan03,Phleps06,Zheng07,Abbas10}, and perhaps even more importantly has allowed measurements at several epochs to be combined with the evolution of the halo properties to understand the evolution of galaxy properties in a cosmological context \citep{Yan03,Conroy06,Zheng07,White07,Wake08a,Brown08,Conroy09,Abbas10}.   

At $z > 1$ the picture becomes less clear, mainly as a result of the difficulty in constructing complete volume limited samples of galaxies at these early epochs. The most precise clustering measurements have come from samples of Lyman Beak Galaxies (LBGs). These galaxies show strong clustering strengths which depends on their luminosity \citep{Adelberger05a,Adelberger05b,Ouchi05,Lee06,Lee09,Hildebrandt09,Bielby10}. However, these samples comprise of relatively blue, unobscured, star-forming galaxies and do not represent a complete sample. In particular, the LBG selection misses the most massive galaxies, which tend to be red and faint in the optical and require deep near-infrared imaging for their selection \citep{Dokkum06}.

Several studies of the clustering of $z > 1$ massive galaxies selected using a variety of optical/near-infrared color selection techniques have been undertaken: Extremely Red Objects \citep[EROs;][]{Daddi00,Roche02,Brown05,Kong06,Kong09,Kim10}, BzKs \citep{Kong06,Hayashi07,Blanc08,Hartley08,McCracken10}, and Distant Red Galaxies \citep[DRGs][]{Grazian06,Foucaud07,Quadri07,Quadri08,Kim10}. These studies revealed strong clustering and some limited evidence for a luminosity and color dependence. However, due to the relatively poor quality of the photometric redshifts of these samples and the effect the color selection has in limiting the range of galaxy types selected, it has been difficult to draw any strong conclusions regarding the relationship between luminosity or stellar mass to halo mass at these redshifts.

This situation is beginning to change with the advent of wide-field near-infrared cameras, which have enabled the construction of wide and deep near infra-red selected galaxy samples at $z > 1$. Whilst it is still almost impossible to generate complete spectroscopic samples of galaxies at these redshifts, it has been possible to combine multiple near-IR bands with deep optical imaging to produce reasonable photometric redshifts and stellar mass estimates. For example, \citet{Foucaud10} combine near-IR imaging from the Palomar Observatory Wide-Field Infrared Survey with optical imaging from the CFHT to define galaxy samples selected by redshift and stellar mass at $z < 2$, based on photometric redshifts accurate to $\delta z/(1+z) \simeq 0.07$. They then use these samples to measure the stellar mass dependent clustering and by using a simple halo model relate, the galaxy stellar mass to the dark matter halo mass.

In this work we make similar measurements using the NEWFIRM medium band survey \citep[NMBS;][]{Dokkum09}. The NMBS combines deep near-IR imaging through five medium band filters, with multiple deep optical, ultra-violet and IR band imaging to produce precise ($\delta z/(1+z) \lesssim 0.02$) photometric redshifts and stellar mass estimates. We use these data to measure the clustering as a function of stellar mass for complete stellar mass limited samples with masses $>7\times10^9\Msun$ and $1 < z < 2$. We then use the latest halo modeling techniques to relate the stellar mass of galaxies to the mass of the halos in which they reside. 

In Section \ref{sec:data} we describe the NMBS. In Section \ref{sec:2pt} we describe how we define the stellar mass limited samples and the calculation of the correlation function. In Section \ref{sec:clusSM} we present our measurements of the clustering as a function of stellar mass. We describe the halo model in Section \ref{sec:halo} and the resulting relationships between stellar mass and halo mass in Section \ref{sec:SMHM} and summarize and conclude in Section \ref{sec:conclusions}.

Throughout this paper, we assume a flat $\Lambda$--dominated CDM cosmology with $\Omega_m=0.27$, $H_0=73 $km s$^{-1} $Mpc$^{-1}$, and $\sigma_8=0.8$ unless otherwise stated.

\section{Data}
\label{sec:data}
\subsection{The NEWFIRM Medium Band Survey}
\label{sec:NMBS}

\begin{figure*}

\vspace{8.0cm}
\includegraphics{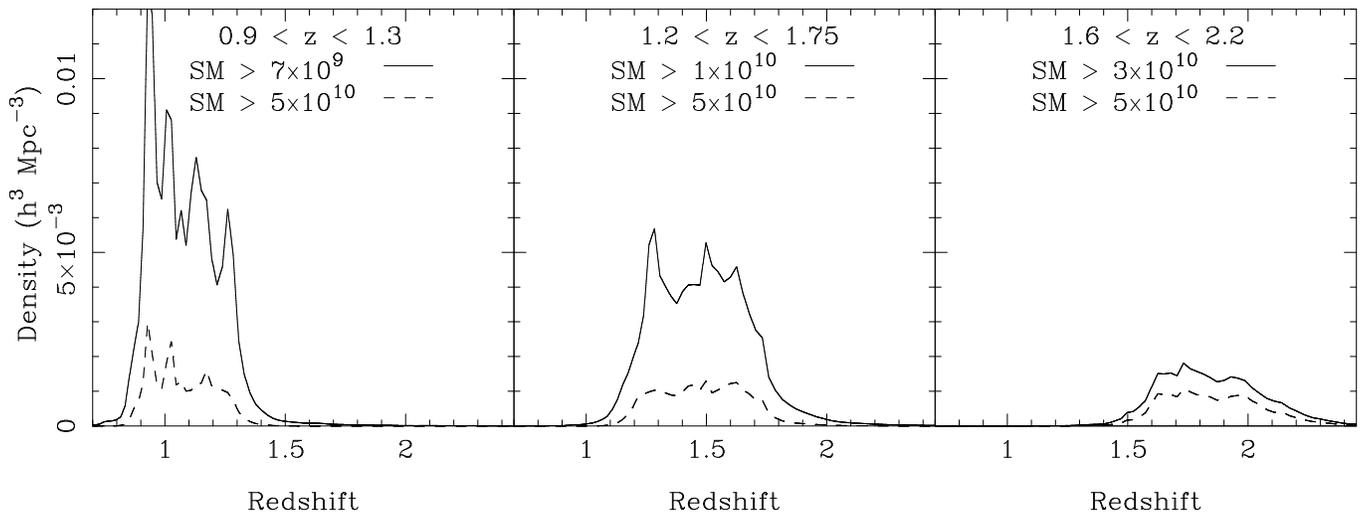}
\caption{\small Example of redshift distributions in the three redshift intervals calculated by summing the photometric redshift PDFs of all galaxies with each galaxy PDF weighted by the fraction of the probability lying within the redshift interval in question (see text for details). For each redshift interval we show the redshift distribution for lowest stellar mass limited sample (solid lines) and for stellar masses $> 5\times10^{10} \Msun$ (dashed lines). Large scale structures are clearly visible in the redshift distributions reflecting the accuracy of the photometric redshifts. 
\label{fig:nz}}
\end{figure*}

The galaxy samples are selected from the NMBS, a moderately wide, moderately 
deep near-infrared imaging survey \citep{Dokkum09}. The survey used the 
NEWFIRM camera on the Kitt Peak 4m telescope. The camera
images a $28' \times 28'$ field with four arrays.  The gaps between the arrays 
are relatively small, making the camera very effective for deep imaging of
$0.25$ deg$^2$ fields. We developed a custom filter system for
NEWFIRM, comprised of five medium bandwidth filters in the wavelength
range 1\,$\mu$m -- 1.7\,$\mu$m. As shown in
\citet{Dokkum09}, these filters pinpoint the Balmer and 4000\,\AA\
breaks of galaxies at $1.5<z<3.5$, providing accurate photometric
redshifts and improved stellar population parameters.  The survey targeted 
two $28'\times 28'$ fields: a subsection of the COSMOS field 
\citep{Scoville07} and a field containing part of the AEGIS strip \citep{Davis07}.
Coordinates and other information are given in \citet{Dokkum09}. Both fields have
excellent supporting data, including ultraviolet (GALEX), extremely deep optical $ugriz$  
(CFHT Legacy Survey\footnote{http://www.cfht.hawaii.edu/Science/CFHLS/})
and deep mid-IR (Spitzer IRAC and MIPS~\citep{Barmby06,Sanders07}) imaging. 
The reduced CFHT mosaics were kindly provided to us by the CARS team \citep{Erben09, Hildebrandt09}.
Additionally, the COSMOS fields includes deep Subaru $BVriz$ and 12 optical
medium-band images\footnote{http://irsa.ipac.caltech.edu/data/COSMOS/images/}.  
The NMBS adds six filters: \Ja, \Jb, \Jc, \Ha, \Hb, and $K$.
Filter characteristics of the five medium band filters are given in \citet{Dokkum09},
and the AB zero points can be found in \citet{Whitaker10b}.

The data reduction, analysis, and properties of the catalogs are
described in \citet{Whitaker10b}. In the present study
we use a $K$-selected catalog based on the full NMBS data set \citep[as described in ][]{Whitaker10b}.
All optical and near-IR images were convolved to the same point-spread
function (PSF) before measuring aperture photometry.
Following previous studies \citep{Labbe03,Quadri07} photometry
was performed in relatively small ``color'' apertures to optimize
the S/N ratio. Total magnitudes in each band
were determined from the SExtractor AUTO aperture flux~\citep{Bertin96}, 
with an additional aperture correction computed from the $K$
band growth curve. The aperture correction is a point-source based correction
that accounts for flux outside of the AUTO aperture.  
We note that about 10\% of the objects detected by SExtractor are classified as single objects but
are actually blended.  We use a deblended catalog here and refer the reader to
\citet{Whitaker10b} for the details of the deblending algorithm employed.

Photometric redshifts were determined with the EAZY code
\citep{Brammer08}, using the full NUV--8$\mu$m spectral energy distributions (SEDs) (NUV--$K$ for objects
in the $\sim 50$\,\% of our AEGIS field that does not have Spitzer coverage).
Publicly available redshifts in the COSMOS and AEGIS fields indicate that the redshift errors are very small
at $\sigma_z/(1+z)<0.02$ \citep[see][]{Brammer09, Whitaker10, Whitaker10b}. Although
there are very few spectroscopic redshifts of optically-faint $K$-selected galaxies in these fields, we note that
we found a similarly small scatter in a pilot program targeting
galaxies from the \citet{Kriek08} near-IR spectroscopic sample \citep[see][]{Dokkum09}.  

Each galaxy within the survey is assigned a weight in each of the NMBS bands based on
the fraction of the maximum exposure present at the galaxies position. In order to ensure both a minimum and even S/N coverage for our samples, we only use galaxies in areas that have a minimum weight ($w_{min}$) $>$ 0.3, i.e. the least exposed optical to near IR band has at least 30\% of the exposure time as the most exposed part of the image.
The regions around bright stars are also excluded from our analysis, as faint galaxies in these regions are either obscured by the foreground star or have systematically incorrect magnitudes due to scattered light.
After the removal of these regions the total remaining area is 0.201 deg$^2$ in the AEGIS field and 0.189 deg$^2$ in COSMOS.

Stellar masses and other stellar population parameters were determined with
FAST \citep{Kriek09}, using the models of \citet{Maraston05},
the \citet{Calzetti00} reddening law, and exponentially declining star formation histories.
Masses and star formation rates are based on a \citet{Kroupa01} initial mass function (IMF);
following \citet{Brammer08} rest-frame near-IR wavelengths are
down-weighted in the fit as their interpretation is uncertain \citep[see, e.g.,][]{Wel06}.
More details are provided in \citet{Brammer09}.

There exist significant systematic uncertainties in the determination of galaxy stellar masses based on uncertainties in the IMF, stellar population synthesis model, extinction law and star formation history which can result in uncertainties in the mass of a factor of a few \citep[e.g. see ][]{Marchesini09,Muzzin09,Conroy09}. In this work our main interest is in determining the dependence of the clustering amplitude on stellar mass and then characterizing that dependence in terms of halo mass. What we are really interested in is not the absolute determination of the stellar mass but the rank order, as we wish to find all galaxies above some stellar mass limit. Whilst the systematics may cause some scatter in this order it is much less than the systematic error on the overall normalization.  

\section{The 2pt-Correlation Function}
\label{sec:2pt}

\subsection{Stellar Mass Limited Samples}
\label{sec:samp}

The aim of this paper is to investigate how the clustering of galaxies at $1 < z < 2$ depends on their stellar masses, and, with the use of the halo model, determine the relationship between stellar mass and halo mass.
The simplest, most robust and systematic free approach is to define volume limited samples, with a variable stellar mass limit but constant volume. With such samples we may directly compare the angular correlation functions of different stellar mass limited samples, and make use of the simplest implementation of the halo model.

In order to study redshift evolution we first define three galaxy samples with slightly overlapping redshift intervals, $0.9 < z < 1.3$, $1.2 < z < 1.75$ and $1.6 < z < 2.2$. Within each redshift range we define several mass limited samples where the lowest stellar mass limit is defined by the stellar mass limit of the survey at the highest redshift in the redshift bin. The highest stellar mass limit is defined so that there are sufficient galaxies to make a reasonable measurement of the correlation function. These samples are described in Table \ref{tab:samp}. 

We estimate the stellar mass completeness limit at a given redshift as follows. We rank order all of our galaxies by redshift. At the redshift of interest we select the next 1000 galaxies of lower redshift. For those 1000 galaxies we find the 90th percentile in the K-band mass-to-light ratio. Finally the stellar mass completeness is determined as the mass a galaxy would have with this 90th percentile mass-to-light ratio at the 99\% K-band completeness limit of the survey \citep[K = 22.8;][]{Whitaker10b}. For example, for our $0.9 < z < 1.3$ sample we wish to estimate the completeness at $z = 1.3$ and so select the 1000 galaxies lying immediately below that redshift which corresponds to the range $1.258 < z < 1.3$. We then find the 90th percentile of the K-band mass-to-light ratio of these 1000 galaxies. At $K = 22.8$, the completeness limit of our survey, this corresponds to a stellar mass of $7.08 \times 10^{9} \Msun$. Since the galaxies we use in this calculation are at lower redshift than our target ($z = 1.3$) they will be more complete in stellar mass than the actual galaxies at $z = 1.3$ yet are sufficiently close in redshift to be representative of those galaxies.

To determine which galaxies are included in any given redshift interval we make use of the full redshift probability distribution functions (PDFs) output by the EAZY photometric redshift code. All galaxies that have any non-zero part of their PDF within the given redshift interval are included in our samples. However, each is given a weight equal to the fraction of the total probability that is within the interval. This has the effect of down weighting galaxies with uncertain redshifts that lie close to the edge of the desired redshift interval, or those with multiple peaks, where one peak is within and one or more is outside. We use these weights throughout when calculating redshift distributions, space densities and the angular correlation function. Table \ref{tab:samp} lists the total number of galaxies contributing to each sample ($N_{Gal}$), the sum of the weights of these galaxies ($N_D$), which gives the effective number of `true' galaxies in the sample, and the mean redshift of the weighted sample.

Figure \ref{fig:nz} shows the redshift distributions of two samples with different stellar mass limits from each redshift interval, calculated using the PDF weights. Whilst there are many galaxies with some probability outside of the redshift intervals the effect of the weight combined with the accuracy of the photometric redshifts is to cause a relatively sharp transition. Large scale structures are clearly visible in the redshift distributions reflecting the accuracy of the photometric redshifts. The same structure is repeated for all stellar mass limits.

\begin{table*}
  \begin{center}
    \caption{\label{tab:samp} Details of the stellar mass limited samples and power law fits to their correlation functions}
	\begin{tabular}{llcrrcrrcrrc}
	\tableline\tableline
      	\multicolumn{6}{c}{} &
      	\multicolumn{3}{c}{$0.0011 < \theta < 0.11, \gamma = 1.8$} &
      	\multicolumn{3}{c}{$0.01 < \theta < 0.11, \gamma = 1.6$} \\
      	\multicolumn{1}{c}{z range} &
      	\multicolumn{1}{c}{SM} &
      	\multicolumn{1}{c}{$\bar{z}$} &
      	\multicolumn{1}{c}{$N_{Gal}$} &
      	\multicolumn{1}{c}{$N_{D}$} &
      	\multicolumn{1}{c}{$\rho$} &
      	\multicolumn{1}{c}{$A_w$} &
      	\multicolumn{1}{c}{$r_0$} &
      	\multicolumn{1}{c}{$\frac{\chi^2}{dof}$} &
      	\multicolumn{1}{c}{$A_w$} &
      	\multicolumn{1}{c}{$r_0$} &
      	\multicolumn{1}{c}{$\frac{\chi^2}{dof}$}\\
	\tableline

0.9 $< z <$ 1.3 &  0.7  &   1.10 & 8970 & 3756.0 &   6.16 & 5.10 $\pm$ 0.50 & 5.80 $\pm$ 0.32 & 0.20 & 9.60 $\pm$ 2.05 & 5.86 $\pm$ 0.78 & 0.03 \\
0.9 $< z <$ 1.3 &  1.0  &   1.10 & 7068 & 3037.5 &   5.03 & 5.60 $\pm$ 0.50 & 6.07 $\pm$ 0.30 & 0.29 & 9.60 $\pm$ 2.10 & 5.82 $\pm$ 0.80 & 0.03 \\
0.9 $< z <$ 1.3 &  1.9  &   1.10 & 3969 & 1842.0 &   3.13 & 6.80 $\pm$ 0.80 & 6.66 $\pm$ 0.44 & 0.62 & 10.10 $\pm$ 2.75 & 5.90 $\pm$ 1.01 & 0.05 \\
0.9 $< z <$ 1.3 &  3.0  &   1.09 & 2666 & 1320.3 &   2.28 & 7.50 $\pm$ 0.95 & 6.97 $\pm$ 0.49 & 0.69 & 10.70 $\pm$ 3.10 & 6.06 $\pm$ 1.11 & 0.04 \\
0.9 $< z <$ 1.3 &  5.0  &   1.09 & 1361 &  727.4 &   1.27 & 10.40 $\pm$ 1.95 & 8.28 $\pm$ 0.86 & 0.52 & 14.40 $\pm$ 5.75 & 7.23 $\pm$ 1.83 & 0.04 \\
1.2 $< z <$ 1.75 &  1.0  &   1.50 & 7623 & 3814.7 &   3.54 & 4.30 $\pm$ 0.45 & 6.23 $\pm$ 0.36 & 1.53 & 7.20 $\pm$ 1.80 & 5.75 $\pm$ 0.90 & 1.11 \\
1.2 $< z <$ 1.75 &  1.9  &   1.50 & 4629 & 2358.4 &   2.24 & 5.60 $\pm$ 0.70 & 7.14 $\pm$ 0.50 & 1.14 & 9.60 $\pm$ 2.35 & 6.79 $\pm$ 1.05 & 0.90 \\
1.2 $< z <$ 1.75 &  3.0  &   1.50 & 3264 & 1682.4 &   1.62 & 5.50 $\pm$ 0.65 & 7.00 $\pm$ 0.46 & 2.02 & 10.50 $\pm$ 2.15 & 7.11 $\pm$ 0.91 & 0.99 \\
1.2 $< z <$ 1.75 &  5.0  &   1.50 & 1739 &  938.5 &   0.93 & 6.50 $\pm$ 1.00 & 7.55 $\pm$ 0.65 & 1.42 & 13.20 $\pm$ 2.85 & 8.04 $\pm$ 1.09 & 0.50 \\
1.2 $< z <$ 1.75 &  6.0  &   1.49 & 1288 &  709.2 &   0.71 & 6.40 $\pm$ 1.25 & 7.50 $\pm$ 0.81 & 1.61 & 14.50 $\pm$ 3.40 & 8.54 $\pm$ 1.26 & 1.62 \\ 
1.6 $< z <$ 2.2 &  3.0  &   1.87 & 2900 & 1501.2 &   1.09 & 5.40 $\pm$ 0.75 & 7.21 $\pm$ 0.56 & 0.61 & 9.40 $\pm$ 2.25 & 6.82 $\pm$ 1.03 & 0.04 \\
1.6 $< z <$ 2.2 &  5.0  &   1.87 & 1631 &  879.5 &   0.67 & 6.30 $\pm$ 1.15 & 7.66 $\pm$ 0.78 & 0.45 & 11.00 $\pm$ 3.05 & 7.31 $\pm$ 1.28 & 0.26 \\
1.6 $< z <$ 2.2 &  7.0  &   1.88 & 942 &  513.0 &   0.40 & 7.50 $\pm$ 1.30 & 8.30 $\pm$ 0.80 & 0.33 & 15.00 $\pm$ 3.45 & 8.72 $\pm$ 1.26 & 0.20 \\
1.6 $< z <$ 2.2 & 10.0  &   1.87 & 487 &  277.0 &   0.22 & 14.00 $\pm$ 2.75 & 11.49 $\pm$ 1.26 & 0.76 & 22.40 $\pm$ 6.90 & 10.93 $\pm$ 2.12 & 0.17 \\

	\tableline
    \end{tabular}
\tablecomments{Stellar masses (SM) are in units of $10^{10} \Msun$, the space density ($\rho$) has units $10^{-3} h^{3} Mpc^{-3}$. The errors are 1 sigma.}
\end{center}
\end{table*}

The 2pt-correlation function is a straight forward way to measure the spatial clustering of our galaxy samples, and when combined with the space density can produce very strong constraints on the distribution of galaxies within dark matter halos. Since we do not have sufficiently precise redshift information for our sources, we choose to calculate the angular correlation function and then relate it to the real space correlation function using Limber's equations \citep{Limber54}.

\subsection{The angular correlation function}

The 2pt-angular correlation function, $w(\theta)$, is defined as the excess probability above Poisson of finding an object at an angular separation $\theta$ from another object. This is calculated by comparing the number of pairs as a function of angular scale in our galaxy catalogs, with the number in a random catalog, which covers the same angular region as our data. We make this measurement using the \citet{Landy93} estimator,
\begin{equation}
	w(\theta) = \frac{1}{RR(\theta)}\left[DD(\theta)\left(\frac{n_R}{n_D}\right)^2 - 2DR(\theta)\left(\frac{n_R}{n_D}\right) + RR(\theta)\right]
\end{equation}
where $DD(\theta)$, $DR(\theta)$ and $RR(\theta)$ are data-data, data-random and random-random pair counts respectively, and $n_D$ and $n_R$ are number of galaxies in the data and random catalogs.  

We generate random catalogs for each galaxy sample following the angular masks of the survey, i.e. excluding areas with W$_{min} <$ 0.3 and those around bright stars. The random catalog has a constant space density and at least 20 times the number of random points as data points. 

As discussed in Section \ref{sec:samp} we associate a weight with each galaxy calculated as the fraction of the photometric redshift PDF that lies within a given redshift slice. When calculating the correlation function each pair is weighted as the multiple of the weights of each galaxy in the pair and the pair count is then the sum of the weights over all pairs in the angular bin. Similarly the normalization factor $n_D$ is the sum of the weights of all galaxies in the sample and is given in Table \ref{tab:samp}. This means that galaxies that are most likely to lie within our desired redshift interval are given more weight in the correlation function calculation, which should lead to a higher signal to noise measurement. This scheme is in essence similar to the one proposed by \citet{Myers09}. 

Our weighting is equivalent to a Monte Carlo approach of randomly assigning a redshift to each galaxy based on its PDF, applying the redshift cut, calculating the correlation function, repeating many times and finally calculating the mean of all these correlation functions. We verify this by applying this procedure to one of our galaxy samples calculating the correlation function 100 times and find the mean correlation function agrees with the weighed correlation function to within 1\% on all scales. We find a variance of $<$10\% on all scales which indicates the expected error due to the photometric redshifts if one were to use a single sample that utilized the best photometric redshift estimate. The weighting scheme should yield errors due to the photometric redshift uncertainties somewhat smaller than this.

We note that if we just use the best photometric redshift rather the PDF, selecting all galaxies that have a best photometric redshift within a given redshift range and using no weight, the resulting angular correlation functions are almost identical to those calculated with the PDF weighting. This is consistent with the error estimated from the Monte-Carlo approach.  This again reflects the accuracy of our photometric redshifts since most of the galaxies selected in this manner have all of their redshift PDF within the redshift range.

The calculated angular correlation functions for the lowest stellar mass limited samples in all three redshift ranges are shown in Figure \ref{fig:fields}. They show the characteristic power law shape, with evidence of a break at $\sim 1 Mpc$ becoming more apparent as higher redshift as expected by the halo model (see Section \ref{sec:halo}).  

\subsection{Integral constraint}

Since our fields cover a relatively small area, we expect the integral constraint \citep{Groth77} to have a significant effect on our clustering measurements, leading to the underestimation of the clustering strength by a constant factor ($IC$). $IC$ is equal to the fractional variance of the galaxy counts on the size of the field. Therefore, the magnitude of $IC$ depends on both the field size and the clustering strength of the sample, increasing as the field size decrease and the clustering increases.

Following \citet{Infante94} and \citet{Roche99} we numerically estimate IC using

\begin{equation}
\label{eq:IC}
	IC = \frac{\sum_i w(\theta_i)RR(\theta_i)}{\sum_i RR(\theta_i)}.
\end{equation}

This estimate is often made using an iterative process: A model for $w(\theta_i)$ is fit to the observed $w(\theta)$ and $IC$ is calculated using equation \ref{eq:IC}. This correction is then applied to the observed $w(\theta)$, the model re-fit, and $IC$ recalculated until there is convergence. It is typical to assume a power-law as the functional form of $w(\theta)$, however, unless the slope is fixed this iterative process tends to produce large values of $IC$ and flat slopes. We therefore choose to determine the integral constraint when fitting the halo model (see Section \ref{sec:halo}). The halo model almost exactly reproduces the shape of the correlation function, something a simple power-law does not. During the fitting process $IC$ is calculated for each model correlation function and subtracted from the model before being fit to the data. As a test of our procedure we also estimated $IC$ using the mock galaxy catalogs we generated from the Millennium simulation to calculate the errors on the correlation function (see Section \ref{sec:err}). The difference between the correlation function calculated for the full simulation and the mean of the correlation functions in the multiple sub-regions the size of our fields give a direct measurement of IC. We find that both methods give consistent results with values ranging from 0.0150 $\pm$ 0.0005 to 0.033 $\pm$ 0.001.

\subsection{The Real Space Correlation Function}

Given a redshift distribution, the angular correlation function can be directly determined from the real space correlation function, $\xi(r)$, using Limber's equation \citep{Limber54} 

\begin{equation}
\label{eq:limber}
	w(\theta) = \frac{2}{c}\int_0^\infty dz H(z)\left(\frac{dn}{dz}\right)^2 \int_0^\infty du~\xi(r = \sqrt{u^2 + x^2 (\bar{z})\theta^2})
\end{equation}

where $c$ is the speed of light, $H(z)$, the Hubble constant at redshift z, is given by $H(z) = H_0 \sqrt{\Omega_M (1+z)^3 + \Omega_\Lambda}$ assuming a flat universe, $dn/dz$ is the normalized redshift distribution, and $x(\bar{z})$ is the comoving distance to the median redshift.  

Conversely, if a functional form is assumed for the real space correlation function then an accurate measurement of both the angular correlation function and the redshift distribution can be used to determine \xir. For the case where \xir\ is a power law,

\begin{equation}
	\xi(r) = \left(\frac{r}{r_0}\right)^\gamma
\end{equation}

then \wt\ is also a power law

\begin{equation}
\label{eq:plw}
	w(\theta) = A_w \theta^\delta
\end{equation}

where $\delta = \gamma - 1$ and 

\begin{equation}
\label{eq:limberr0}
	A_w = \frac{r_0^\gamma}{c} \Gamma(1/2)\frac{\Gamma\left(\frac{\gamma-1}{2}\right)}{\Gamma\left(\frac{\gamma}{2}\right)} \int_0^\infty dz~H(z) \left(\frac{dn}{dz}\right)^2 x(z)
\end{equation}

with $\Gamma$ indicating the gamma function. We make use of Equation \ref{eq:limber} when fitting the halo model in Section \ref{sec:halo} and Equation \ref{eq:limberr0} when comparing clustering amplitudes in Section \ref{sec:clusSM}.

In both Equations \ref{eq:limber} and \ref{eq:limberr0} $dn/dz$ is the normalized redshift distribution for a given sample without any clustering, i.e. it reflects the selection function of the galaxy sample. Large scale structure is clearly visible in our redshift distributions and so we remove its effects by making a polynomial fit to each redshift distribution. We note that despite the presence of visible structure using the fit rather than the measured $dn/dz$ makes essentially no difference to any of our results. Even though our photo-z errors are small enough to reveal large scale structure the $rms$ error still corresponds to $\simeq60 h^{-1}Mpc$, sufficiently smearing out the clustering signal on large enough scales such that it has a negligible effect.

\subsection{Estimating Errors}
\label{sec:err}

\begin{figure*}

\vspace{8.0cm}
\includegraphics{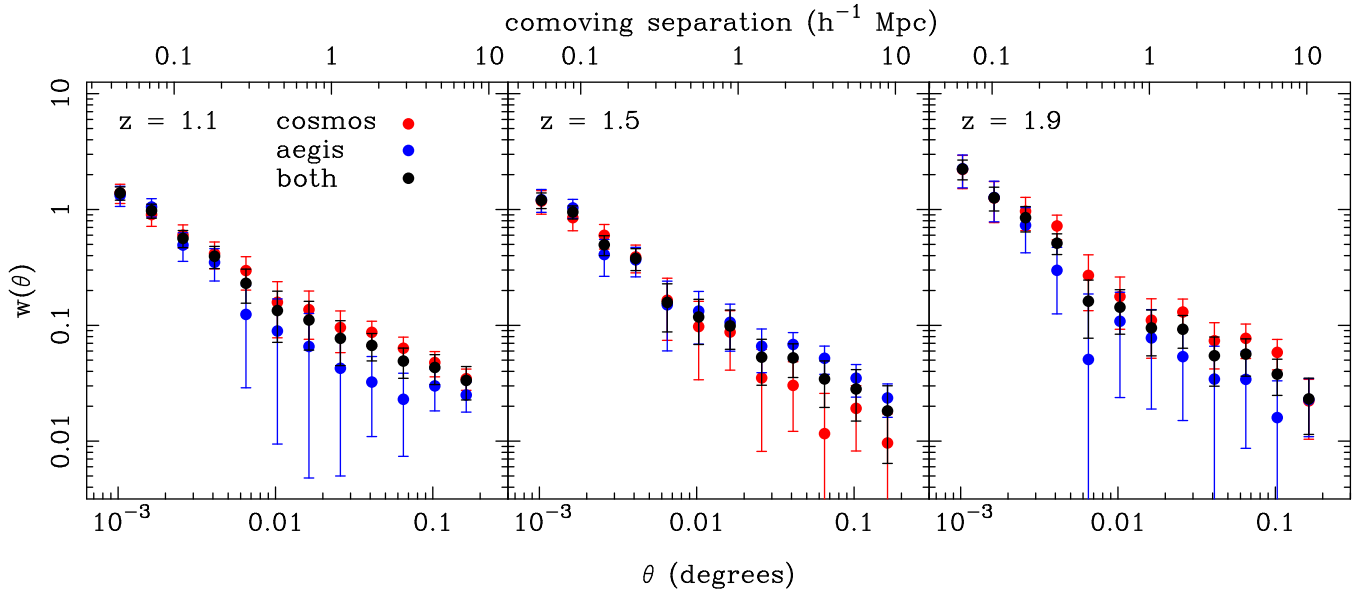}
\caption{\small The angular correlation function for the AEGIS field (blue), the COSMOS field (red) and both fields combined (black) for the lowest mass limit samples in each redshift interval. The top axis shows the angular scale converted to a comoving separation at the mean redshift of each sample. All the correlation functions show the typical power law form and do show some evidence of a break at around 0.01 degrees, particularly in the highest redshift bin. There is also evidence of cosmic variance between the two fields at a level consistent with the magnitude of the errors and given the covariance on large scales. 
\label{fig:fields}}
\end{figure*}

Whether we wish to make comparisons between the clustering of our samples or fit models to the clustering we need to make an accurate estimate of the measurement errors and the correlation between the data points in the form of a covariance matrix. 

There are a number of ways in which the errors on clustering measurements maybe estimated; simple Poisson errors, internal estimates such as jackknife and bootstrap re-sampling and mock galaxy catalogs based on N-body simulations or analytical halo distributions. Poisson errors are known to be an underestimate of the true error, particularly on large scales, and do not provide an estimate of the covariance. 

Both jackknife and bootstrap re-sampling methods have been shown to produce reasonable estimates of the full covariance \citep[e.g.][]{Zehavi05} but may suffer from some systematic issues \citep{Norberg09}. They are also only really effective where the scales of interest are significantly smaller than the region of the survey being removed in each re-sampling. 

Mock catalogs, in which simulated galaxies have the same clustering properties as the real galaxies, may provide the best error estimates since large numbers of individual surveys can be created and the full covariance between them accurately calculated. The only potential problem with this method is that there could be higher order clustering effects not encoded in the 2-point correlation function from which the mocks are generated.

We have chosen to use mock catalogs generated from the Millennium simulation \citep{Springel05} to generate the covariance matrices we use for all fits to the measured correlation functions. We discuss in Appendix \ref{sec:sims} the details of our approach, and the reasons for this choice. In brief, we first fit a Halo Occupation Distribution (HOD; see Section \ref{sec:halo}) to the correlation function and space density of the observed NMBS samples, using an estimate of the errors from jackknife re-sampling. We then populate halos according to this HOD in the full Millennium simulation box at the appropriate redshift for the sample. The Millennium box is then split in to multiple sub-regions with the same geometry as the survey fields and the correlation function is calculated for each. The covariance matrix is then generated from the correlation functions of each sub-region. 

We note that when comparing the clustering between our stellar mass limited samples these errors may well be an overestimation since we assume that each measurement is independent, when in fact it is made within the same volume. They are the correct errors if one wished to compare with a similar measurement made in a different region of sky. We discuss this issue further in Appendix \ref{sec:sims}.

\subsection{Uncertainties in the redshift PDFs}

As described above, the photometric redshift PDFs generated with the EAZY code are used twice in our analysis; first, to assign a weight to each galaxy in the correlation function calculation, and second to calculate the redshift distribution which is used to convert the angular to the real space correlation function. Although these PDFs are calibrated against the thousands of spectroscopic redshifts in our fields, the relative lack of spectroscopic coverage in the $1.2 < z < 2.2$ interval of interest to this paper means we cannot be absolutely certain of their reliability in this range. 
That being said, we do have good reason to expect any uncertainty in the PDFs to have a small effect on our measurements. Only galaxies that have some part of their PDF outside of the redshift range of the sample are being affected by the use of the PDFs in our \wt~calculations. Since the redshift intervals that we select are much broader than the PDFs, the number of these galaxies should be small, consisting of those lying close to the redshift boundaries or those with multiple probability peaks that are widely separated. 

In order to explicitly test the effect of uncertainties in the PDFs, we modify the calibration to generate two new sets of PDFs that are approximately either half or twice the width of the best estimate of the true PDF. While we are confident that our calibration is more accurate than this factor of two, these new samples will allow us to determine the effect in a worst-case scenario.

We rerun the sample selection, clustering calculations and halo model fits (see Section \ref{sec:halo}) for all of the samples using both new sets of PDFs. We find that, at most, the clustering amplitude is changed by 4\%. As expected, this occurs for the samples with the lowest stellar mass limit at each redshift, thus the faintest galaxies, where the PDFs are already broadest. When we fit the halo model, which includes the effect of the modified PDFs on the redshift distribution and the space density as well as the clustering, we find that the HOD parameters are changed by at most 6\%, again for the lowest mass limit. It is interesting to note that there is an even smaller effect on the derived parameters such as the bias, mean halo mass and satellite fraction, which change by no more than 2\%. 

Since these modified PDFs represent something of a worst-case scenario, and any differences are significantly less than the measurement errors, we are confident that any inaccuracies in the PDFs are not a concern for our analysis. 

It is important to note that while the above test demonstrates that our analysis is relatively insensitive to the photometric redshift PDF accuracy, within reasonable limits, it does not mean that we could produce similar quality measurements of the dependence of clustering on stellar mass with less accurate photometric redshifts. A reduction in the redshift accuracy would increase the error on the stellar mass estimates, causing our mass limits to be poorly defined. This would cause the samples we define to become increasing less close to being volume limited, making the standard halo model assumptions invalid, and the fraction of catastrophic failures would increase, causing systematic variations in the redshift and stellar mass distributions. 

\subsection{Field to Field Variation}

Figure \ref{fig:fields} shows the angular correlation function of the lowest mass limited sample in the three redshift intervals in both the AEGIS and COSMOS fields separately and for both fields combined.  The errors plotted in these correlation functions, and in any further correlation function plots, are the square root of the diagonal terms of the covariance matrices generated using the mock catalogs. A correction due to the integral constraint calculated from the HOD model has been applied. 

Visually there are differences between the correlation functions from the two fields: the COSMOS field shows stronger clustering at $\bar{z} = 1.1$ on large scales, while the opposite is true at $\bar{z} = 1.5$. Within these redshift bins the same trends continue for the higher mass limited samples. This variation is to be expected, as cosmic variance will play a role for fields of this size. However, it should be noted that the differences in the correlation functions between the fields are only of marginal significance, particularly when the full covariance is taken into account, due to the highly correlated nature of the largest scale measurements. 

In the remainder of the paper we will consider only measurements from the combined fields, reducing both the statistical and cosmic variance errors.

\section{Clustering as a Function of Stellar Mass}
\label{sec:clusSM}

The goal of this paper is to investigate how the clustering of galaxies depends on galaxy stellar mass, and thus determine the relationship between stellar mass and dark matter halo mass at redshift $1 < z < 2$. In this Section we present our 2-point correlation function measurements for the mass limited samples and then test to see if there is a significant dependence on stellar mass.

\begin{figure*}
\vspace{13.5cm}
\includegraphics{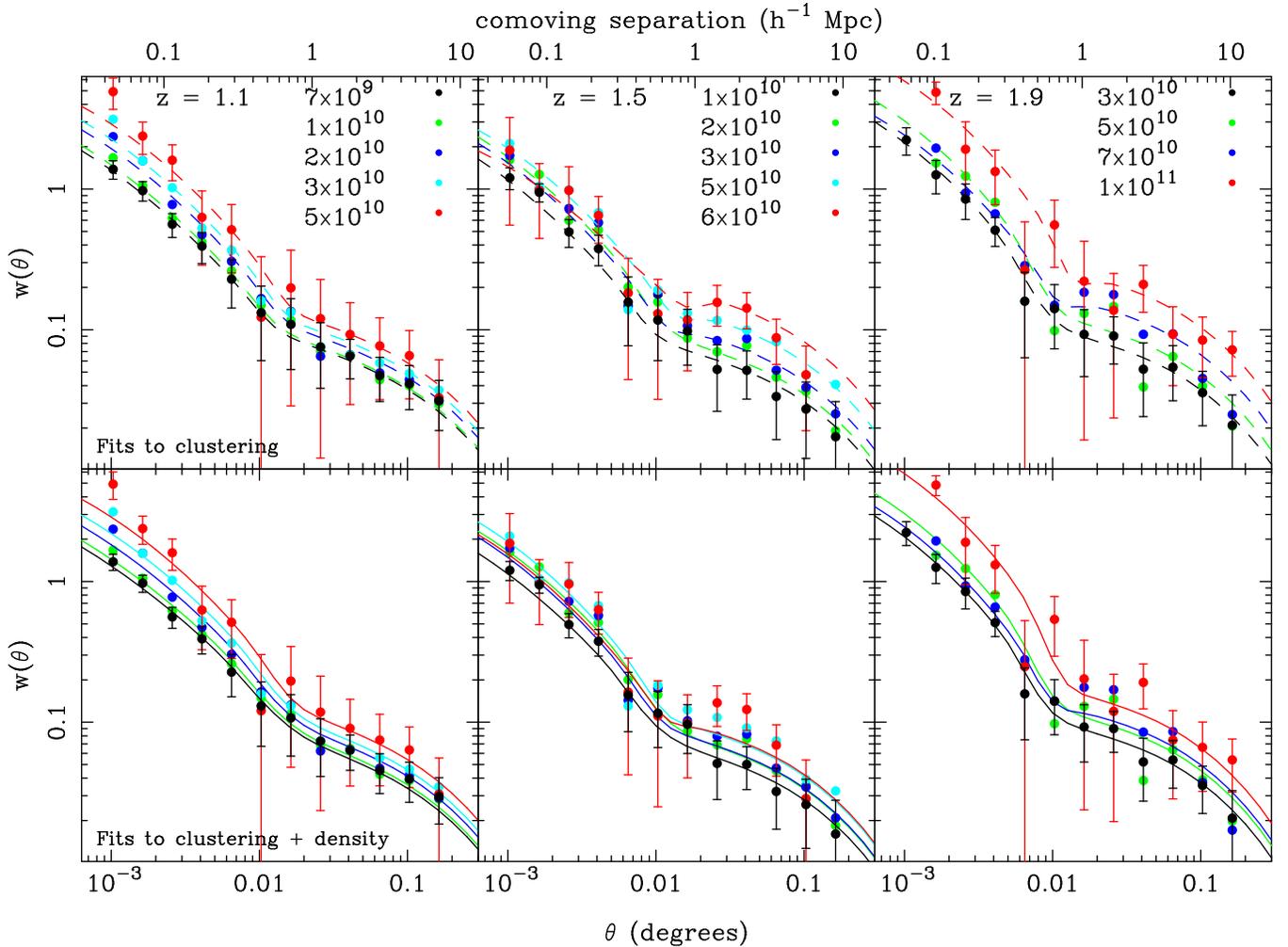}
\caption{\small The angular correlation function as a function of stellar mass limit in each redshift interval. The the stellar mass limits in units of $\Msun$ are given in the legend. For clarity error bars are only shown for the lowest and highest stellar mass limits. The top axis shows the angular scale converted to a comoving separation at the mean redshift of each sample. The solid lines (bottom panel) show the best fitting HOD model fit to both the clustering and density, whereas the dashed lines (top panel) are fits to the clustering only (see Section \ref{sec:halo} for details). In each redshift interval the clustering amplitude increases as the stellar mass increases. The HOD fits show a similar increase in amplitude with stellar mass. When the space density is included in the fits the clustering amplitude at large scales is reduced; this is particularly noticeable at $z = 1.5$ but is present at all redshifts (see Section \ref{sec:tension}). 
\label{fig:2ptSM}}
\end{figure*}

Figure \ref{fig:2ptSM} shows the 2-point correlation functions for all of the stellar mass limited samples in the three redshift intervals. A trend is visible on all scales and at all redshifts of an increasing clustering amplitude with increasing stellar mass limit. 

When considering angular correlation functions, it is always important to remember that the amplitude depends on both the intrinsic clustering of the population ($\xi(r)$) and the normalized redshift distribution (equation \ref{eq:limber}). However, our samples are defined to be very close to being volume limited, resulting in the normalized redshift distributions being very similar for all of the mass limits within a single redshift interval. It is therefore reasonable to make a direct comparison between the angular clustering measurements in this case. 

To quantify the significance of the stellar mass dependent clustering, we calculate the $\chi^2$ between the correlation functions for the lowest and highest stellar mass samples in each redshift bin using the covariance matrices for both samples. We choose to fit in the range $0.0011 < \theta < 0.11$ degrees. The lower limit is chosen to ensure we are not being affected by any remaining deblending issues; the upper limit represents where the data become very poorly measured and highly correlated with the smaller scale points. 

This $\chi^2$ test shows that in all three redshift intervals there is a significant difference between the high and low stellar mass correlation functions. The most significant difference is for the $\bar{z}$ = 1.1 sample where we find $\chi^2$ = 22.87 with 10 degrees of freedom ({\it dof}) and a probability of 1.1\% of the high and low stellar mass clustering measurements being the same. At $\bar{z}$ = 1.5 we measure a $\chi^2$ = 21.01 with 10 {\it dof} and a probability of 2.1\% and at $\bar{z}$ = 1.9 we measure a $\chi^2$ = 18.75 with 10 {\it dof} and a probability of 4.3\%. While the highest redshift sample has the lowest significance, showing a trend with stellar mass at 95\%, it also has the smallest stellar mass range, a factor of 3.33 compared to factors of 7 and 6 for the $\bar{z}$ = 1.1 and $\bar{z}$ = 1.5 respectively.

As we have already discussed, differences in the redshift distributions between our mass limited samples could result in differences in the amplitude of the clustering that are not intrinsic. Even though we do not believe that this is a significant issue for these data it is reassuring to make tests that are independent of this effect. In Section \ref{sec:halo} the halo model fits will be free from this issue since the redshift distributions are used in the halo model calculations. But first we shall use a simpler approach, which has long been used in the literature, of modeling the 2-point correlation function as a power-law. 

Initially we fit a power-law to \wt~(equation \ref{eq:plw}) leaving both the normalization ($A_w$) and slope ($\delta$) as free parameters. We choose to fit both over the full angular range of the correlation function ($0.0011 < \theta < 0.11$ degrees) and over a restricted range ($0.011 < \theta < 0.11$ degrees) which just covers the large scale clustering. Within the halo model there are two terms that contribute to the overall correlation function: the 1-halo term from pairs within halos, and the 2-halo term from pairs between halos. This results in a characteristic feature at the scale of the typical halo size and can often result in a change of slope at this scale \citep{Berlind02}. This transition is well established in measurements of the correlation function \citep[e.g.][]{Zehavi04} and can be seen in the correlation functions shown in Figure \ref{fig:2ptSM}. 

Since we want to investigate how the amplitude of the clustering varies with stellar mass we must fix the slope and just fit for the amplitude. We find that slopes ($\delta$) of 0.8 for all scales and 0.6 for large scales are consistent with the individual best fits for all of the samples and so fit with the slopes fixed to these values. Using the best fitting amplitudes, the redshift distributions and equation \ref{eq:limberr0}, we can calculate the correlation length of the real space correlation function ($r_0$). 

\begin{figure}

\vspace{12.5cm}
\includegraphics{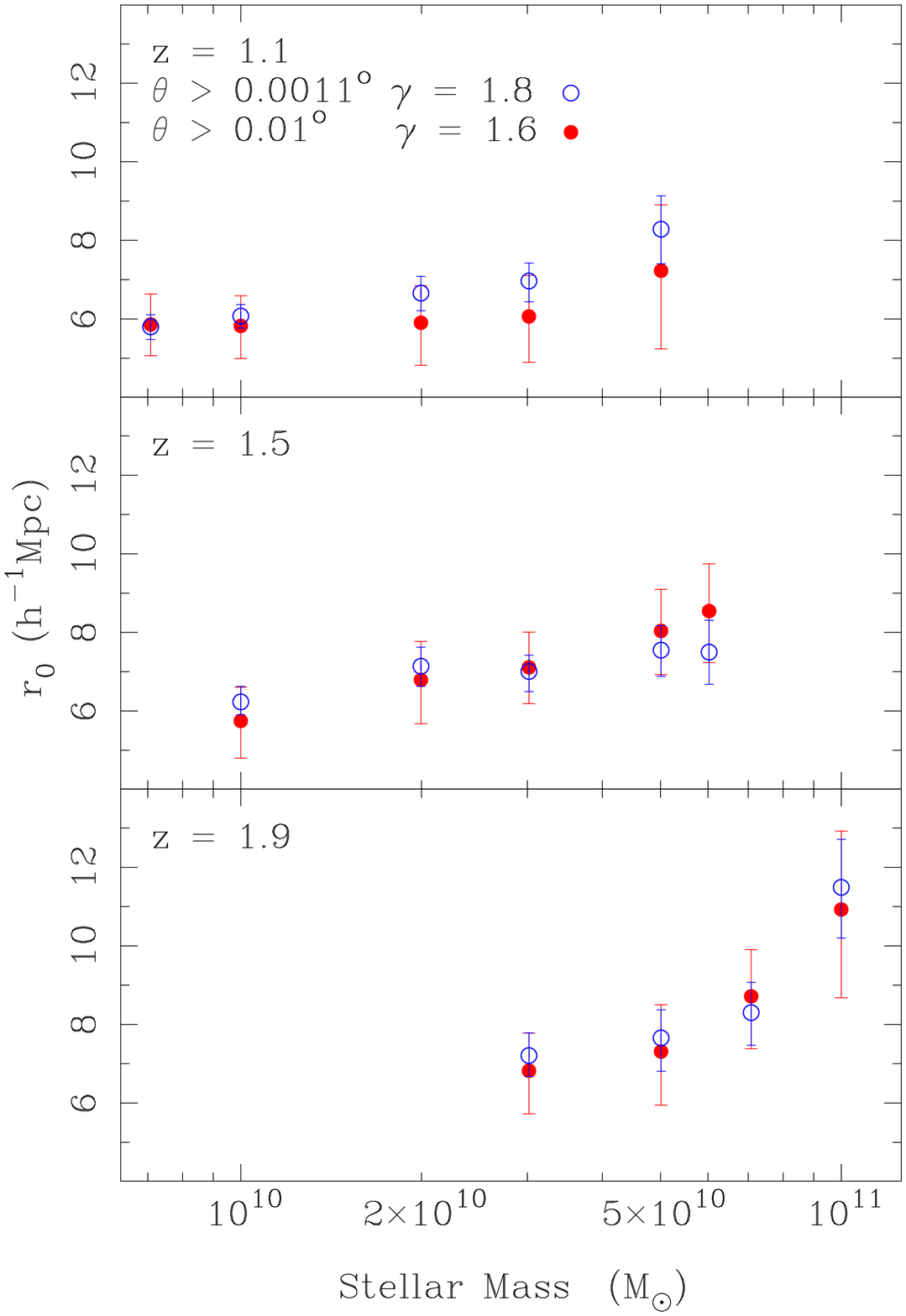}
\caption{\small The spatial correlation length, $r_0$, as a function of stellar mass limit in the three redshift ranges. Values of $r_0$ determined from power law fits to the full correlation function with the slope fixed at 1.8 (open circles) and from fits to just the large scales with the slope fixed at 1.6 (filled circles) are shown. As the stellar mass increases the clustering amplitude in real space also increases.
\label{fig:r0sm}}
\end{figure}

Figure \ref{fig:r0sm} shows the best fitting values of $r_0$ as a function of stellar mass for the three redshift intervals (see Table \ref{tab:samp} for the values). Significant trends of an increasing correlation length with stellar mass limit are present in all three redshift intervals and for the fits to both angular ranges, with the exception of the large scale fit at $\bar{z} = 1.1$ where the trend is not very significant. This confirms the results of the direct \wt~comparisons above, showing that the clustering strength does depend significantly on stellar mass at $1 < z < 2$.

\section{Halo Model Analysis}
\label{sec:halo}

\begin{figure}
\vspace{8.5cm}
\includegraphics{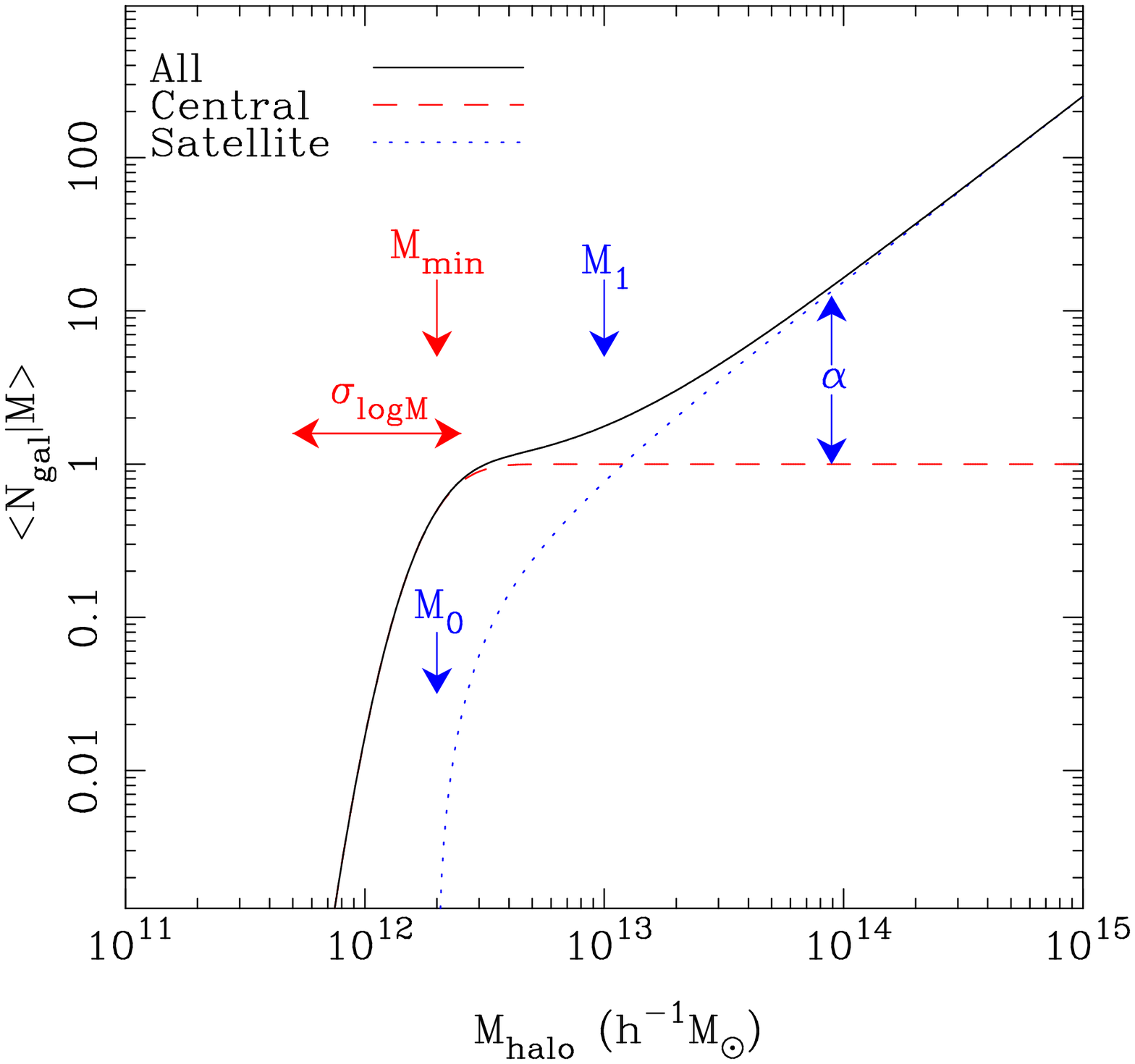}
\caption{\small An example HOD showing the mean number of galaxies per halo as a function of halo mass. The dashed line shows the central galaxy distribution, the dotted line the satellite distribution and the solid line the total. Values of the mass scales $\Mmin$, $\M1$ and $M_0$ are indicated by the arrows. The affects of $\sigM$ and $\alpha$ are also indicated. In brief $\Mmin$ and $\M1$ are the mass thresholds for central and satellite galaxies respectively. $\sigM$ controls the rate of truncation of the central galaxy distribution. $\alpha$ is the slope of the satellite power law distribution, and $M_0$ is the cut off mass for the satellite power law.  
\label{fig:NMexample}}
\end{figure}

The halo model (see Cooray \& Sheth 2002 for a review) assumes that the galaxy clustering signal encodes information about the Halo Occupation Distribution (HOD) - how the galaxies populate Dark Matter halos - in particular, how the HOD depends on halo mass. In essence, the HOD describes the probability distribution that a halo of a given mass ($M$) host a certain number ($N$) of galaxies of given type, $P(N|M)$. 

In the halo model, every galaxy is associated with a halo and all halos are 200 times the background density whatever the mass $M$ of the halo.  Sufficiently massive halos typically host more than one galaxy.  The halo model we use distinguishes between the central galaxy in a halo, and the other galaxies, which are usually called satellites.  This approach is motivated by simulations \citep[e.g.][]{Kravtsov04,Zheng05} and has been a standard assumption of semi-analytic galaxy formation models for many years \citep[e.g.][]{Baugh06}.  There is now strong observational evidence that these two types of galaxies are indeed rather different, and that the halo model parameterization of this difference is accurate \citep{Skibba07}.  

When considering a sample of galaxies with a fixed luminosity or stellar mass limit, an HOD that is close to a step function for central galaxies and a power law for satellites is a reasonable approximation. We choose to use the parameterization introduced by \citet{Zheng07} which has a soft cut off in the central galaxy HOD, allowing for the scatter in the stellar mass halo mass relation, and a cut in the satellite power law at low halo mass. This five parameter analytic HOD was motivated by the desire to match the HODs from the simulations presented in \citet{Zheng05} and has since been shown to precisely reproduce the measured clustering in luminosity limited samples, which favor both the varying soft central cut and the satellite cut \citep{Zheng07,Brown08,Blake08,Ross09,Ross10,Zehavi10}. Our clustering measurements are not sufficiently precise to accurately constrain all five parameters simultaneously, and so we choose to fix several of them in our analysis (see \ref{sec:HODfits}). However, we choose to keep the five parameter form to enable an easier comparison to other work at lower redshifts.

The details of the halo model calculation are similar to those presented in \citet{Wake08a,Wake08b}, however there are significant differences which we describe here. We present a full description of our halo model calculation in Appendix \ref{sec:AppHalo}. 

Following \citet{Zheng07}, the fraction of halos of mass $M$ which host centrals is modeled as 
\begin{equation}
\label{eq:Ncen}
	\langle N_c|M\rangle = \frac{1}{2}\left[1 + \erf\left(\frac{logM - logM_{min}}{\sigma_{logM}}\right)\right]. 
\end{equation}

Only halos which host a central may host satellites.  In such halos, the number of satellites is drawn from a Poisson distribution with mean 
\begin{equation}
 \label{eq:Nsat}
	\langle N_s|M \rangle = \left(\frac{M - M_0}{M_1^\prime}\right)^{\alpha}.
\end{equation}
Thus, the mean number of galaxies in halos of mass $M$ is 
\begin{equation}
 \label{eq:Ntot}
 \langle N|M\rangle = \langle N_c|M\rangle[1 + \langle N_s|M \rangle], 
\end{equation} 
and the predicted number density of galaxies is 
\begin{equation}
\label{eq:den}
	n_g =  \int dM\, n(M)\, \langle N|M\rangle,
\end{equation}
where $n(M)$ is the halo mass function, for which we use the latest parameterization given by \citet{Tinker10b}.

We further assume that the satellite galaxies in a halo trace an NFW profile \citep{Navarro96} around the halo center, and that the halos are biased tracers of the dark matter distribution.  The halo bias ($b(M)$)depends on halo mass in a way that can be estimated directly from the halo mass function \citep{Sheth99}, and we use the most up to date parameterization of \citet{Tinker10b}. 

In \citet{Wake08a,Wake08b} we used the linear theory power spectrum ($P_{Lin}(k)$) throughout the calculation, whereas we now use the non-linear power spectrum at the redshift of interest when calculating the 2-halo term. We also apply the scale dependent bias and halo exclusion corrections given by \citet{Tinker05}.

With these assumptions the halo model for $\xi(r)$ is completely specified \citep[e.g.][]{Cooray02}.  We then calculate $w(\theta)$ from $\xi(r)$ using equation~(\ref{eq:limber}). 

In addition to $\xi(r)$, we are interested in the satellite fraction, 
\begin{equation}
\label{eq:Fsat}
 F_{sat} = \int dM\, n(M)\,\langle N_c|M\rangle\,\langle N_s|M\rangle/n_g,  
\end{equation} 
the fraction of the galaxies in a given sample that are satellite galaxies in halos,
and two measures of the typical masses of galaxy host halos: 
an effective halo mass 
\begin{equation}
\label{eq:Meff}
	M_{eff} =  \int dM\, M\, n(M)\, \langle N|M\rangle/n_g,
\end{equation}
and the average effective bias factor 
\begin{equation}
\label{eq:blin}
	b_{g} =  \int dM\, n(M)\, b(M)\,\langle N|M\rangle/n_g,
\end{equation}
where $b(M)$ is the halo bias.

We show in Figure \ref{fig:NMexample} an example HOD where we indicate the effect of each of the five parameters in our model. $\Mmin$ is the mass threshold for central galaxies, and represents the halo mass which hosts on average 0.5 galaxies above the stellar mass limit. $\sigM$ determines the cutoff profile for the central galaxies with higher values corresponding to a more gentle cutoff. Both $\Mmin$ and $\sigM$ also have more physical meaning (see \citet{Zheng05} and \citet{Zehavi10} for details). In brief, the value of $\Mmin$ for a sample of galaxies with a stellar mass limit $SM_{min}$, corresponds to the halo mass that hosts central galaxies with a median stellar mass of $SM_{min}$. $\sigM$ is proportional to the scatter in the stellar mass of galaxies living in halos of mass $\Mmin$.
The halo occupation of satellite galaxies are described by the characteristic mass $\M1$, the slope of the power law $\alpha$ and the low mass cut off $M_0$.

\begin{table*}
  \begin{center}
    \caption{\label{tab:HODfitc} HOD and derived parameters from fits to the clustering only. }
	\begin{tabular}{lcccccccccc}
	\tableline\tableline
      	\multicolumn{1}{c}{$\bar{z}$} &
      	\multicolumn{1}{c}{SM} &
      	\multicolumn{1}{c}{$M_{min}$} &
      	\multicolumn{1}{c}{$M_1^\prime$} &
      	\multicolumn{1}{c}{$M_1$} &
      	\multicolumn{1}{c}{$n_g$} &
      	\multicolumn{1}{c}{$b_g$} &
      	\multicolumn{1}{c}{$F_{sat}$} &
      	\multicolumn{1}{c}{$M_{eff}$} &
      	\multicolumn{1}{c}{$\frac{\chi^2}{dof}$} &
      	\multicolumn{1}{c}{Fit prob} \\
	\tableline
1.1 &  0.7 & 0.17$^{+0.09}_{-0.06}$ & 1.18$^{+1.37}_{-0.67}$ & 1.32$^{+1.44}_{-0.69}$ & 2.51$^{+2.26}_{-1.10}$ & 2.17$^{+0.13}_{-0.08}$ & 0.22$^{+0.10}_{-0.07}$ & 0.90$^{+0.10}_{-0.05}$ & 0.52 & 0.860\\ 
1.1 &  1.0 & 0.17$^{+0.08}_{-0.06}$ & 1.08$^{+1.18}_{-0.61}$ & 1.20$^{+1.31}_{-0.63}$ & 2.49$^{+2.18}_{-1.05}$ & 2.20$^{+0.11}_{-0.08}$ & 0.24$^{+0.11}_{-0.08}$ & 0.95$^{+0.08}_{-0.05}$ & 1.10 & 0.355\\ 
1.1 &  2.0 & 0.26$^{+0.13}_{-0.08}$ & 1.69$^{+1.95}_{-0.81}$ & 1.91$^{+2.08}_{-0.81}$ & 1.47$^{+1.01}_{-0.66}$ & 2.37$^{+0.16}_{-0.10}$ & 0.21$^{+0.08}_{-0.07}$ & 1.14$^{+0.17}_{-0.10}$ & 1.78 & 0.066\\ 
1.1 &  3.0 & 0.30$^{+0.29}_{-0.11}$ & 1.84$^{+4.92}_{-1.10}$ & 2.09$^{+5.50}_{-1.18}$ & 1.23$^{+1.32}_{-0.78}$ & 2.44$^{+0.30}_{-0.12}$ & 0.21$^{+0.12}_{-0.11}$ & 1.24$^{+0.37}_{-0.13}$ & 0.77 & 0.643\\ 
1.1 &  5.0 & 0.38$^{+0.66}_{-0.16}$ & 2.07$^{+12.51}_{-1.33}$ & 2.51$^{+13.34}_{-1.60}$ & 0.92$^{+1.29}_{-0.72}$ & 2.59$^{+0.56}_{-0.18}$ & 0.22$^{+0.14}_{-0.15}$ & 1.45$^{+0.86}_{-0.21}$ & 1.12 & 0.343\\ 
1.5 &  1.0 & 0.15$^{+0.11}_{-0.07}$ & 0.96$^{+1.91}_{-0.70}$ & 1.10$^{+1.92}_{-0.77}$ & 2.08$^{+3.78}_{-1.19}$ & 2.60$^{+0.25}_{-0.16}$ & 0.20$^{+0.17}_{-0.09}$ & 0.61$^{+0.14}_{-0.06}$ & 1.14 & 0.330\\ 
1.5 &  2.0 & 0.29$^{+0.47}_{-0.08}$ & 2.13$^{+11.19}_{-1.05}$ & 2.51$^{+1.12}_{-1.19}$ & 0.83$^{+2.57}_{-0.53}$ & 2.96$^{+0.62}_{-0.24}$ & 0.15$^{+0.17}_{-0.08}$ & 0.86$^{+0.09}_{-0.08}$ & 1.13 & 0.339\\ 
1.5 &  3.0 & 0.38$^{+0.49}_{-0.28}$ & 4.21$^{+17.20}_{-3.89}$ & 4.79$^{+18.12}_{-4.35}$ & 0.54$^{+3.70}_{-0.40}$ & 3.10$^{+0.69}_{-0.54}$ & 0.09$^{+0.27}_{-0.06}$ & 0.95$^{+0.69}_{-0.33}$ & 1.49 & 0.146\\ 
1.5 &  5.0 & 0.67$^{+0.43}_{-0.47}$ & 10.23$^{+19.41}_{-9.35}$ & 10.96$^{+19.23}_{-9.87}$ & 0.22$^{+1.35}_{-0.13}$ & 3.55$^{+0.51}_{-0.75}$ & 0.06$^{+0.19}_{-0.03}$ & 1.37$^{+0.59}_{-0.61}$ & 0.86 & 0.556\\ 
1.5 &  6.0 & 1.17$^{+0.70}_{-0.71}$ & 44.83$^{+88.99}_{-38.82}$ & 47.86$^{+83.96}_{-41.55}$ & 0.08$^{+0.32}_{-0.05}$ & 4.11$^{+0.63}_{-0.90}$ & 0.02$^{+0.06}_{-0.01}$ & 2.04$^{+0.91}_{-0.99}$ & 0.70 & 0.708\\ 
1.9 &  3.0 & 0.17$^{+0.17}_{-0.08}$ & 0.64$^{+2.41}_{-0.49}$ & 0.83$^{+2.48}_{-0.60}$ & 1.36$^{+3.47}_{-0.97}$ & 3.30$^{+0.45}_{-0.23}$ & 0.24$^{+0.22}_{-0.15}$ & 0.55$^{+0.20}_{-0.08}$ & 0.31 & 0.972\\ 
1.9 &  5.0 & 0.26$^{+0.33}_{-0.14}$ & 1.11$^{+6.27}_{-0.92}$ & 1.32$^{+7.00}_{-1.02}$ & 0.67$^{+2.50}_{-0.52}$ & 3.61$^{+0.70}_{-0.37}$ & 0.20$^{+0.27}_{-0.14}$ & 0.69$^{+0.40}_{-0.14}$ & 0.76 & 0.656\\ 
1.9 &  7.1 & 0.56$^{+0.42}_{-0.34}$ & 8.31$^{+24.07}_{-7.32}$ & 9.12$^{+23.99}_{-7.92}$ & 0.16$^{+0.71}_{-0.11}$ & 4.23$^{+0.71}_{-0.77}$ & 0.05$^{+0.15}_{-0.03}$ & 1.03$^{+0.54}_{-0.42}$ & 0.58 & 0.811\\ 
1.9 & 10.0 & 1.17$^{+0.65}_{-0.92}$ & 16.90$^{+35.07}_{-16.50}$ & 17.38$^{+35.10}_{-16.75}$ & 0.04$^{+0.93}_{-0.02}$ & 5.24$^{+0.75}_{-1.47}$ & 0.04$^{+0.36}_{-0.02}$ & 1.84$^{+0.75}_{-1.03}$ & 1.64 & 0.097\\

	\tableline
    \end{tabular}
\tablecomments{$M_{min}$ and $M_1^\prime$ are the fitted HOD parameters. $M_1$, $n_g$, $b_g$, $F_{sat}$ and $M_{eff}$ are all derived parameters and are respectively the mass scale at which a halo hosts 1 satellite on average, the mean galaxy number density, the average linear bias, the satellite fraction, and the effective halo mass. The stellar mass (SM) is in units of $10^{10} \Msun$, halo masses are in units of $10^{13}$\hMsun and density has units $10^{-3} h^{3} Mpc^{-3}$. The errors are 1 sigma marginalized over the other parameters.}
\end{center}
\end{table*}

\begin{table*}
  \begin{center}
    \caption{\label{tab:HODfitd} HOD and derived parameters from fits to the clustering and density. }
	\begin{tabular}{lcccccccccc}
	\tableline\tableline
      	\multicolumn{1}{c}{$\bar{z}$} &
      	\multicolumn{1}{c}{SM} &
      	\multicolumn{1}{c}{$M_{min}$} &
      	\multicolumn{1}{c}{$M_1^\prime$} &
      	\multicolumn{1}{c}{$M_1$} &
      	\multicolumn{1}{c}{$n_g$} &
      	\multicolumn{1}{c}{$b_g$} &
      	\multicolumn{1}{c}{$F_{sat}$} &
      	\multicolumn{1}{c}{$M_{eff}$} &
      	\multicolumn{1}{c}{$\frac{\chi^2}{dof}$} &
      	\multicolumn{1}{c}{Fit prob} \\
	\tableline

1.1 &  0.7 & 0.10$^{+0.01}_{-0.01}$ & 0.42$^{+0.08}_{-0.08}$ & 0.52$^{+0.11}_{-0.09}$ & 5.67$^{+0.92}_{-0.72}$ & 2.08$^{+0.04}_{-0.03}$ & 0.35$^{+0.04}_{-0.03}$ & 0.88$^{+0.05}_{-0.04}$ & 0.69 & 0.735\\ 
1.1 &  1.0 & 0.11$^{+0.01}_{-0.01}$ & 0.47$^{+0.11}_{-0.08}$ & 0.58$^{+0.12}_{-0.10}$ & 4.67$^{+0.70}_{-0.64}$ & 2.14$^{+0.03}_{-0.04}$ & 0.35$^{+0.03}_{-0.04}$ & 0.95$^{+0.03}_{-0.05}$ & 1.18 & 0.295\\ 
1.1 &  2.0 & 0.17$^{+0.01}_{-0.02}$ & 0.76$^{+0.15}_{-0.14}$ & 0.91$^{+0.18}_{-0.15}$ & 2.82$^{+0.54}_{-0.34}$ & 2.25$^{+0.04}_{-0.05}$ & 0.30$^{+0.04}_{-0.03}$ & 1.06$^{+0.06}_{-0.07}$ & 1.91 & 0.039\\ 
1.1 &  3.0 & 0.20$^{+0.03}_{-0.02}$ & 0.85$^{+0.26}_{-0.14}$ & 1.10$^{+0.22}_{-0.18}$ & 2.23$^{+0.27}_{-0.36}$ & 2.35$^{+0.05}_{-0.06}$ & 0.31$^{+0.03}_{-0.05}$ & 1.18$^{+0.08}_{-0.09}$ & 0.82 & 0.610\\ 
1.1 &  5.0 & 0.31$^{+0.04}_{-0.03}$ & 1.45$^{+0.44}_{-0.31}$ & 1.74$^{+0.55}_{-0.29}$ & 1.26$^{+0.17}_{-0.20}$ & 2.51$^{+0.07}_{-0.06}$ & 0.26$^{+0.05}_{-0.04}$ & 1.37$^{+0.13}_{-0.11}$ & 1.13 & 0.337\\ 
1.5 &  1.0 & 0.11$^{+0.01}_{-0.01}$ & 0.50$^{+0.12}_{-0.08}$ & 0.63$^{+0.13}_{-0.11}$ & 3.46$^{+0.53}_{-0.47}$ & 2.51$^{+0.04}_{-0.04}$ & 0.28$^{+0.03}_{-0.03}$ & 0.58$^{+0.03}_{-0.03}$ & 1.13 & 0.335\\ 
1.5 &  2.0 & 0.16$^{+0.01}_{-0.01}$ & 0.62$^{+0.16}_{-0.12}$ & 0.76$^{+0.15}_{-0.13}$ & 2.25$^{+0.28}_{-0.33}$ & 2.71$^{+0.06}_{-0.05}$ & 0.29$^{+0.04}_{-0.04}$ & 0.72$^{+0.05}_{-0.05}$ & 1.17 & 0.304\\ 
1.5 &  3.0 & 0.19$^{+0.02}_{-0.02}$ & 1.05$^{+0.28}_{-0.20}$ & 1.20$^{+0.38}_{-0.20}$ & 1.61$^{+0.23}_{-0.22}$ & 2.73$^{+0.05}_{-0.05}$ & 0.21$^{+0.03}_{-0.03}$ & 0.71$^{+0.04}_{-0.04}$ & 1.48 & 0.140\\ 
1.5 &  5.0 & 0.27$^{+0.03}_{-0.02}$ & 1.74$^{+0.60}_{-0.37}$ & 2.09$^{+0.42}_{-0.50}$ & 0.92$^{+0.13}_{-0.13}$ & 2.94$^{+0.07}_{-0.07}$ & 0.17$^{+0.03}_{-0.04}$ & 0.85$^{+0.06}_{-0.07}$ & 0.90 & 0.528\\ 
1.5 &  6.0 & 0.32$^{+0.03}_{-0.02}$ & 3.04$^{+1.70}_{-0.78}$ & 3.31$^{+1.94}_{-0.80}$ & 0.70$^{+0.08}_{-0.10}$ & 2.99$^{+0.08}_{-0.07}$ & 0.11$^{+0.03}_{-0.04}$ & 0.86$^{+0.08}_{-0.07}$ & 0.85 & 0.583\\ 
1.9 &  3.0 & 0.19$^{+0.02}_{-0.01}$ & 0.83$^{+0.22}_{-0.14}$ & 1.00$^{+0.20}_{-0.17}$ & 1.10$^{+0.13}_{-0.16}$ & 3.36$^{+0.07}_{-0.06}$ & 0.21$^{+0.03}_{-0.03}$ & 0.56$^{+0.03}_{-0.03}$ & 0.30 & 0.980\\ 
1.9 &  5.0 & 0.26$^{+0.02}_{-0.01}$ & 1.11$^{+0.34}_{-0.23}$ & 1.32$^{+0.42}_{-0.22}$ & 0.67$^{+0.09}_{-0.10}$ & 3.61$^{+0.07}_{-0.08}$ & 0.20$^{+0.04}_{-0.04}$ & 0.69$^{+0.04}_{-0.04}$ & 0.75 & 0.677\\ 
1.9 &  7.1 & 0.34$^{+0.03}_{-0.02}$ & 2.70$^{+0.93}_{-0.63}$ & 3.02$^{+0.96}_{-0.73}$ & 0.40$^{+0.05}_{-0.06}$ & 3.76$^{+0.08}_{-0.07}$ & 0.11$^{+0.03}_{-0.02}$ & 0.75$^{+0.05}_{-0.04}$ & 0.61 & 0.808\\ 
1.9 & 10.0 & 0.49$^{+0.03}_{-0.03}$ & 2.20$^{+0.51}_{-0.46}$ & 2.75$^{+0.56}_{-0.46}$ & 0.22$^{+0.03}_{-0.03}$ & 4.23$^{+0.07}_{-0.07}$ & 0.16$^{+0.03}_{-0.02}$ & 1.06$^{+0.05}_{-0.05}$ & 1.63 & 0.091\\

	\tableline
    \end{tabular}
\tablecomments{$M_{min}$ and $M_1^\prime$ are the fitted HOD parameters. $M_1$, $n_g$, $b_g$, $F_{sat}$ and $M_{eff}$ are all derived parameters and are respectively the mass scale at which a halo hosts 1 satellite on average, the mean galaxy number density, the average linear bias, the satellite fraction, and the effective halo mass. The stellar mass (SM) is in units of $10^{10} \Msun$, halo masses are in units of $10^{13}$\hMsun and density has units $10^{-3} h^{3} Mpc^{-3}$. The errors are 1 sigma marginalized over the other fit parameter.}
\end{center}
\end{table*}

\subsection{HOD Fits}
\label{sec:HODfits}

The HOD defined by equations \ref{eq:Ncen} and \ref{eq:Nsat} contains five free parameters: $\Mmin$ and $\sigM$ for central galaxies and $\M1$, $M_0$ and $\alpha$ for satellite galaxies. Our correlation functions are not sufficiently accurate to precisely constrain all five of these parameters. In previous studies of the HOD \citep[e.g.][]{Zheng07,Brown08,Zehavi10} it has been found that $M_0 \simeq \Mmin$ and we fix this relationship when fitting the HOD. Furthermore, we find that both $\sigM$ and $\alpha$ are very poorly constrained by our measurements. There are reasonably strong theoretical arguments based on the distribution of sub-halos for $\alpha \simeq 1$ for all stellar mass limited samples \citep{Kravtsov04,Zheng05}, which are supported by precise clustering measurements in the local universe \citep[e.g.][]{Zehavi10}. As a result of this and since $\alpha$ has some considerable degeneracy with $\M1$, we choose to fix $\alpha$. We find that $\alpha = 1$ is an acceptable fit ($<1\sigma$) for all samples and so we fix it to this value in the remainder of the analysis. 

There is also some degeneracy between $\Mmin$ and $\sigM$, and since $\sigM$ is so poorly constrained we choose to fix it also. There is substantial observational evidence that $\sigM$ increases with increasing stellar mass \citep{Zheng07,Brown08,Zehavi10}, and we see a hint of this when we allow this parameter to be free. However, the trend is not significant in our measurements, and we find all samples are consistent with $\sigM$ = 0.15 ($<1\sigma$), so we fix it to this value. We are thus interested in measuring how the two mass thresholds for central and satellite galaxies ($\Mmin$ and $\M1$) depend on the stellar mass limit of our samples. 

A given HOD predicts both the clustering and space density of a galaxy population and so both can be used when fitting. The inclusion of the space density provides particularly strong constraints on $\Mmin$ and since it can be expected to be very well measured it is a very useful constraint on the HOD overall. The dominant error on the space density, like the clustering, is cosmic variance; we estimate this error using the same mock catalogs used for the clustering error estimates and find that the typical error on the space density to be about 15\%. 

Tables \ref{tab:HODfitc} and \ref{tab:HODfitd} give the best fitting values of the HOD parameters $\Mmin$ and $\M1$ and their $1\sigma$ errors for fits to just the clustering and the clustering and space density simultaneously for all samples. These Tables also contain the derived parameters $M_1$, the halo mass that on average hosts one satellite galaxy, as well as $n_g$, $F_{sat}$, $M_{eff}$, and $b_g$ as given by equations \ref{eq:den}, \ref{eq:Fsat}, \ref{eq:Meff}, and \ref{eq:blin} respectively. The best fitting model \wt s are shown in Figure \ref{fig:2ptSM}.

\subsubsection{A Discrepancy Between the Clustering and Space Density in the Halo Model?}
\label{sec:tension}

\begin{figure}

\vspace{12.5cm}
\includegraphics{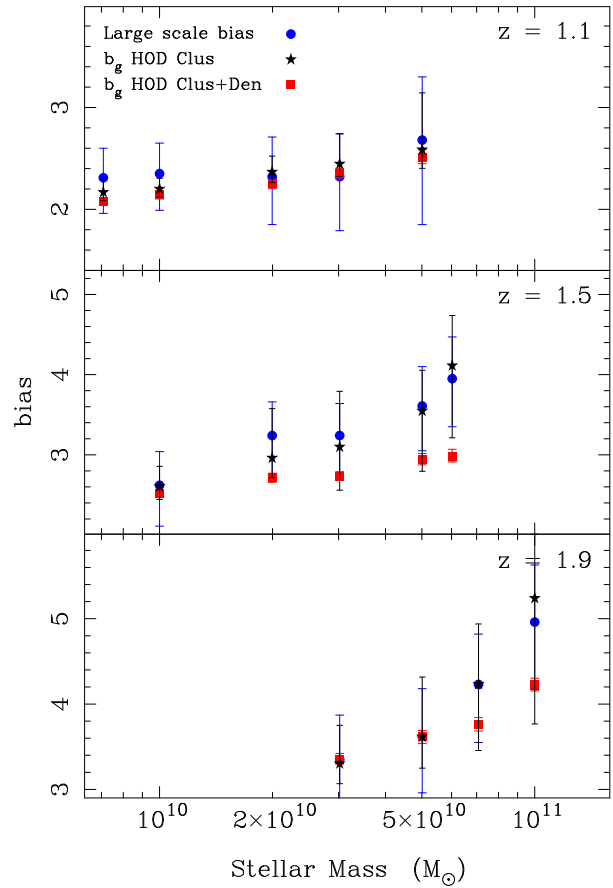}
\caption{\small The galaxy bias as a function of stellar mass limit in the three redshift intervals. The bias is measured in three ways. The blue points are calculated by fitting the non-linear dark matter correlation function to the measured correlation function on scales $>0.024$ degrees. The other two bias measurements are determined from the full HOD fits, the black stars from fits to the clustering and the red squares from fits to both the clustering and space density. There is some evidence of an offset between the bias determined from the clustering and the bias from the fit to the clustering and space density in the sense that the clustering prefers higher bias. This is particularly evident at higher redshift and higher stellar mass limit.
\label{fig:LSbias}}
\end{figure}

\begin{figure}
\vspace{8.5cm}
\includegraphics{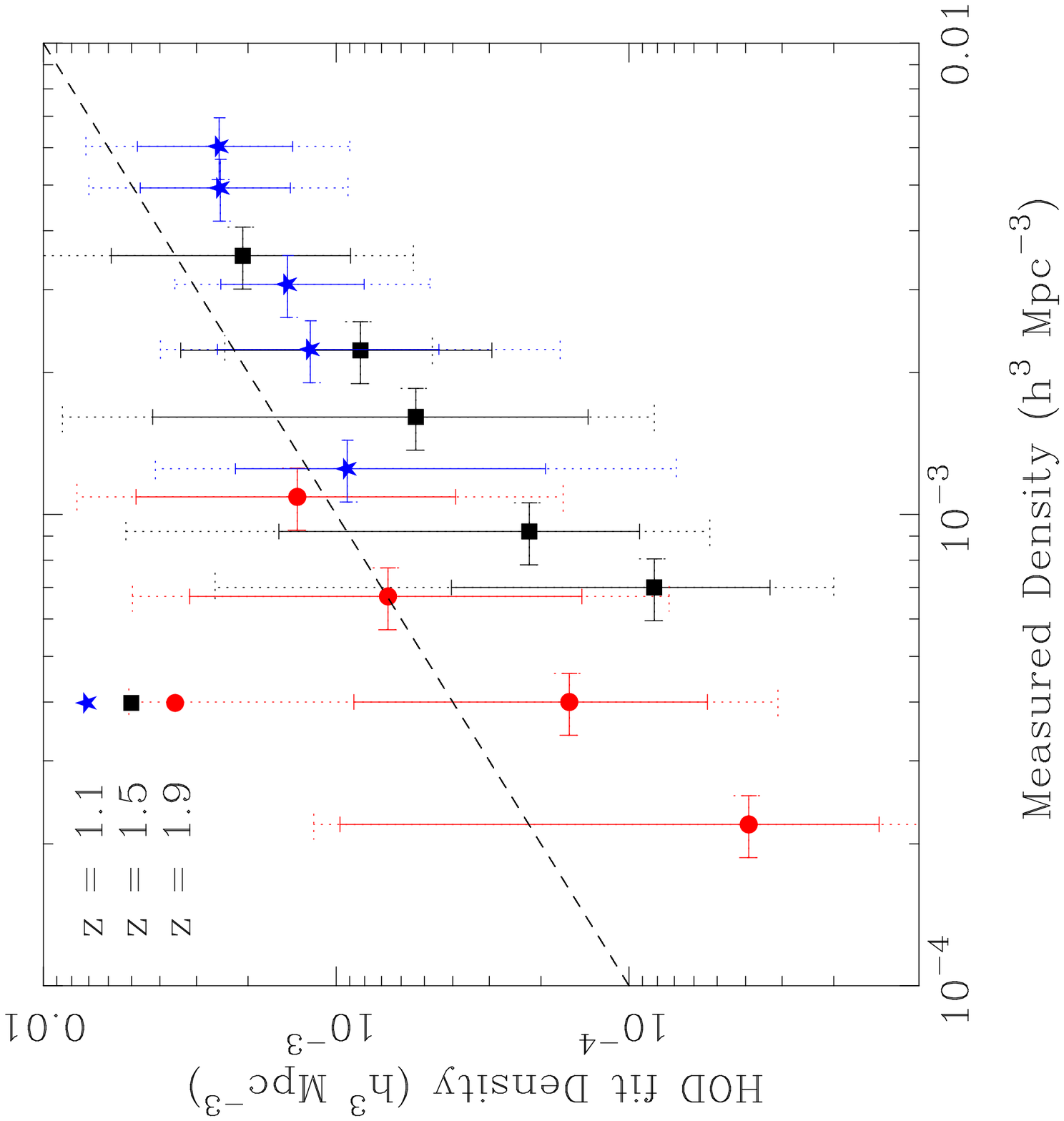}
\caption{\small The predicted mean density based on the HOD fits to the galaxy clustering as a function of the measured space density for all stellar mass limits in the three redshift intervals. The vertical solid and dotted error bars are the 1 and 2 sigma errors on the HOD fit density respectively. The dashed line shows the one to one relation. All but one of the points lie to the right of the one to one relation meaning that the model fit to the clustering implies a lower space density than is measured. 
\label{fig:HODden}}
\end{figure}

If the halo model is a good representation of our data, one would expect the fits based on just the clustering and those based on the clustering and space density to be consistent. Several previous studies of high redshift clustering have reported difficulty in simultaneously fitting both the space density and clustering within the halo model. For example, Quadri et al. (2008;Q08) report this issue for distant red galaxies (DRGs) at $2 < z < 3$ and most recently \citet{Matsuoka10} show a similar discrepancy for the most massive galaxies at $z \lesssim 1$. \citet{Tinker10a} were able to fit the measured DRG clustering and space density from Q08 using a better justified HOD model and the latest halo mass-bias relation from \citet{Tinker10b}, which is steeper at higher bias than the \citet{Sheth01} relation used by Q08. However, they required that the field in Q08 have a higher clustering amplitude than average, but were able to demonstrate using simulations that this was not that unusual due to cosmic variance, with just over 16\% of their simulated surveys showing clustering as strong as observed by Q08.

Figure \ref{fig:2ptSM} and the HOD fit parameters in Tables \ref{tab:HODfitc} and \ref{tab:HODfitd} show a small systematic difference between the observed density and clustering and that which is predicted by the model. This is particularly noticeable for high stellar mass (high halo bias) and at $\bar{z}$ = 1.5 and 1.9. The discrepancy is in the sense that the HOD model requires a smaller space density to match the clustering than is observed. We illustrate this explicitly in two ways in Figures \ref{fig:LSbias} and \ref{fig:HODden}. Figure \ref{fig:LSbias} shows the bias as a function of stellar mass limit determined in three ways. The stars and squares show $b_g$ from the HOD fits as given by Equation \ref{eq:blin} for fits to the clustering and clustering plus space density respectively. The circles show the large scale bias ($b_{ls}$) calculated by fitting the non-linear matter correlation function to the clustering on large scales. Both $b_{ls}$ and $b_{g}$ (fitted to the clustering) show higher values than $b_{g}$ fitted to the clustering and density. A similar trend is seen in Figure \ref{fig:HODden} where we show the predicted mean density based on the HOD fit to the clustering compared to the measured density. All but one of the points lie below the one to one relation, although on both plots all are within 2 sigma of the expected relation.

Allowing the HOD parameters that were fixed to vary does not resolve this discrepancy. If we make the satellite slope $\alpha$ steeper or shallower within reasonable bounds, we find that the satellite mass threshold $\M1$ adjusts to compensate keeping the satellite fraction the same and hardly changing the space density. The same is true for $\Mcut$. Adjusting the softening parameter for the central cut off, $\sigM$, does effect the predicted density for a given clustering amplitude, in the sense that sharper cutoffs, smaller values of $\sigM$, will produce a higher density for the same clustering. Our chosen value of $\sigM$ is already quite low and there is some expectation that it should be higher for our more massive galaxy samples, however even an unphysical instantaneous transition does not reduce the discrepancy by much. 

Therefore, in our analysis we do see some evidence of this discrepancy over two independent fields, and three redshift ranges, even though we are using the latest halo mass function and bias relations from \citet{Tinker05} and \citet{Tinker10b}, as well as the most up to date implementation of the HOD and clustering calculation. However, we should emphasize that the individual HOD fits that include the space density are perfectly acceptable fits and are only slightly worse than the fits to clustering alone. It is still striking that we see that this systematic offset is apparent for nearly all of our samples and particularly those with high bias. The simulations that we conduct to estimate our errors do contain some fields with clustering as strong as we observed for our most massive samples at z = 1.5 and 1.9, but these are rare (Figure \ref{fig:millfit}). Even though cosmic variance could well be playing some part, it is challenging to explain the whole of this systematic trend as cosmic variance, particularly when the very similar findings of \citet{Matsuoka10} at $z \lesssim 1$ are considered. 

One possible explanation is that there may be a deficit of highly biased halos in our model, as a result of either the halo mass function or the halo bias relation that we use. We find that modifying the halo bias relation to make it slightly steeper at high halo masses resolves this problem, although this is a purely arbitrary adjustment. There are several different calibrations of the halo bias relation in the literature \citep[e.g.][]{Sheth99,Sheth01,Hamana01,Seljak04,Wetzel07,Tinker10b}, and it is of course at the high bias end of this relation, where the halos are rarest, and thus the errors largest that we are potentially seeing a difference with the observations. We note that the \citet{Tinker10b} bias relation that we use here minimizes this discrepancy compared to other bias relations in the literature. It is thus possible that further calibration of this relation in the high bias regime may still be needed, along with more precise clustering measurements of sufficiently biased galaxy populations.

The high bias regime is one that has yet to be accurately probed with clustering measurements. At lower redshifts even the most massive galaxies occupy less biased halos. For instance, we could accurately predict the space density for the 5700 brightest Luminous Red Galaxies \citep{Eisenstein01} in the SDSS DR7 \citep{Abazajian09} by fitting our HOD model to their projected 2pt-correlation function calculated as in \citet{Wake08a}. However, even these massive galaxies have a bias of 3, which falls at the lower end of the bias values implied by the clustering of our samples where we see the predicted and observed densities become most discrepant. 

Another possibility lies not with the halos themselves but how the galaxies are placed within the halos. We make the standard assumption that the satellite galaxies trace an NFW profile within each halo. If this were not the case, for instance if the satellites had a broader distribution, it may be possible to place more satellite galaxies in massive halos, thus increasing the large-scale clustering amplitude whilst maintaining the small scale amplitude at a level consistent with observations.

It will be very interesting to see if this problem persists with larger surveys in more fields at these redshifts, such as the forthcoming NMBS-II, or if we have just been very unlucky with cosmic variance in this instance.

Because of this potential issue we will present the remainder of our results for both fits to the clustering alone and to fits combining clustering and space density. The overall trends are the same but there is a stronger stellar mass-to-halo mass dependence for the fits to the clustering alone. However, we will focus mainly on the fits that include the space density. Its inclusion provides a very strong constraint on the HOD, particularly when the small scale clustering is as well measured as it is here. We are comfortable doing this since while there is a hint of a discrepancy within the model it is not yet large enough to exclude the fits including the density at a high significance. 

\section{The stellar mass-to-halo mass relationship}
\label{sec:SMHM}

\begin{figure*}
\vspace{11.5cm}
\includegraphics{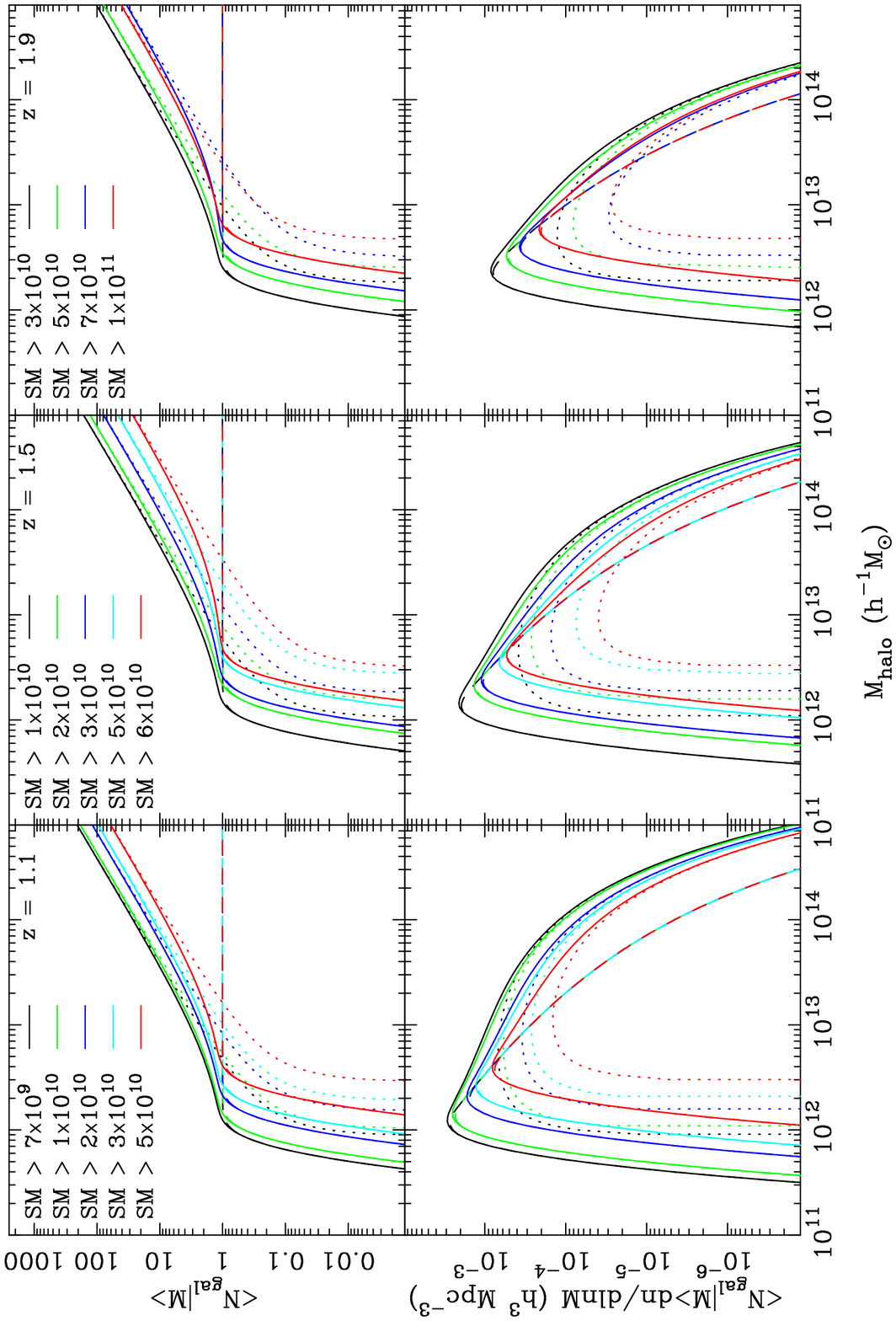}
\caption{\small The mean number of galaxies per halo as a function of halo mass, the HOD, (top) and the mean number of galaxies per halo times the number density of halos as a function of halo mass (bottom). The total, central and satellite contributions are shown by the solid, dashed and dotted lines respectively. In each redshift range the typical halo mass increases as the stellar mass increases.
\label{fig:HODSM}}
\end{figure*}

\begin{figure}
\vspace{12.2cm}
\includegraphics{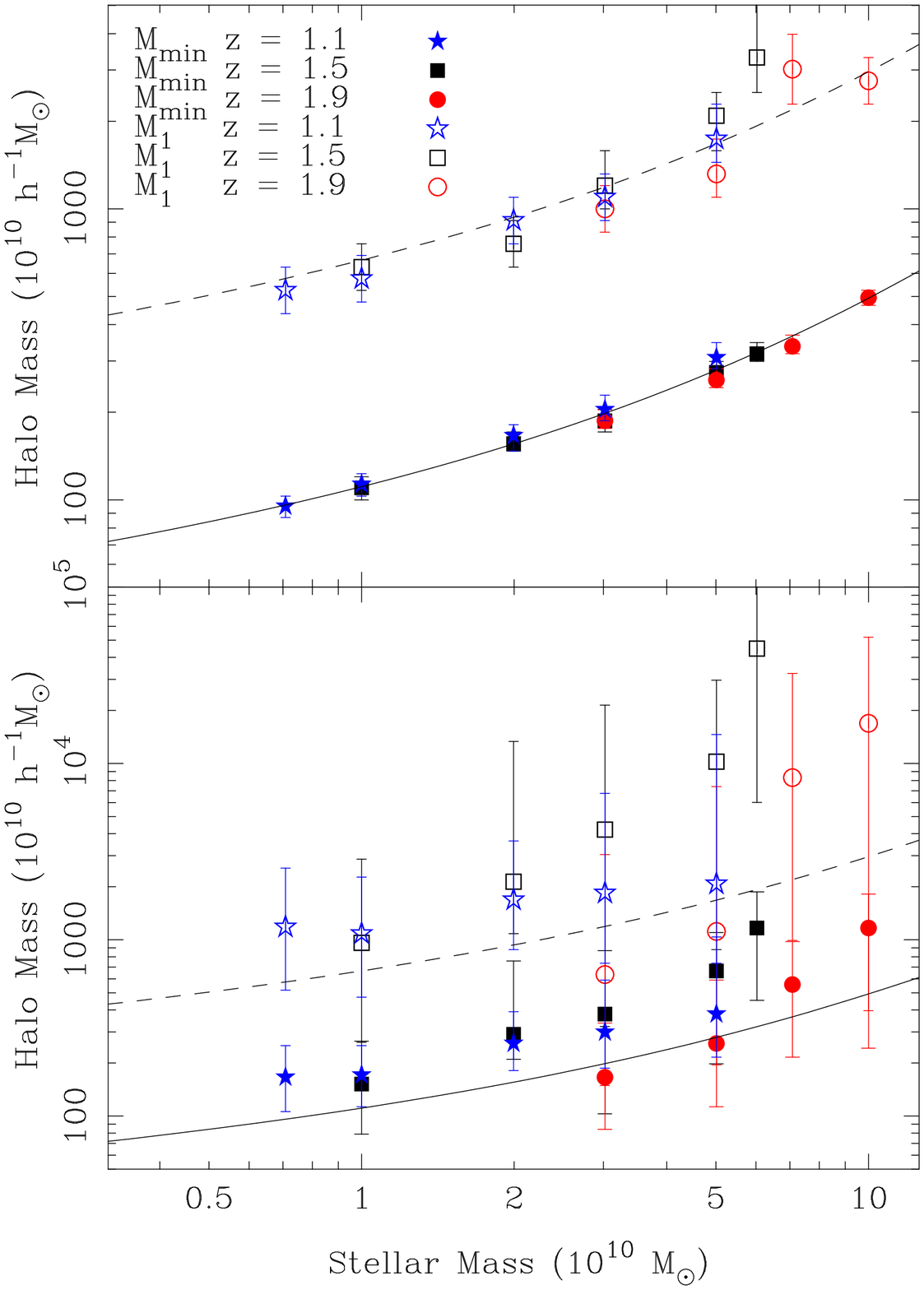}
\caption{\small The relationship between stellar mass and halo mass for the central and satellite galaxies from the HOD fits to the clustering (bottom) and the clustering and density (top). The solid lines in both panels show the best fit stellar mass $M_{min}$ relation of the form given in equation \ref{eq:SMHM} to the clustering and density HOD fits. The dashed lines show the $M_{min}$ relation scaled up by a factor of 6. Higher mass thresholds are typically preferred by the fits to just the clustering than by the fits to both the clustering and density. Both the satellite and central halo mass thresholds increase as the stellar mass limit increases.  
\label{fig:SMHM}}
\end{figure}

Figure \ref{fig:HODSM} shows the HODs (for the fits to the clustering and density), and Figure \ref{fig:SMHM} shows the relationship between the halo mass scale parameters for central and satellite galaxies, $\Mmin$ and $M_1$, as a function of stellar mass limit for fits to the clustering (bottom) and the clustering and space density (top). Once again the clear trend of increasing halo mass with increasing stellar mass is visible at all redshifts, with the effective halo mass increasing by a factors of 2.2 $\pm$ 0.2, 2.5 $\pm$ 0.3, and 5.7 $\pm$ 0.4 per decade in stellar mass at redshifts 1.1, 1.5 and 1.9 respectively. The fits that include the density in Figure \ref{fig:SMHM} show much smaller errors due to the precision of the density measurement and it's constraining power. They also typically prefer lower values of the halo mass thresholds ($\Mmin$ and $\M1$) as discussed in Section \ref{sec:tension}. 

Where there is overlap between the stellar mass bins there is very little evidence for evolution in the HODs with redshift. Only the highest stellar mass bin that is in all three redshift ranges shows significant evolution in $\Mmin$ with redshift, such that $\Mmin$ increases with decreasing redshift. 

\citet{Zheng05} show that for the definition of the HOD for centrals used here (equation \ref{eq:Ncen}), the median stellar mass of central galaxies living in halos of mass $\Mmin$ is the stellar mass limit of the sample. It therefore appears that the median stellar mass of central galaxies at a given halo mass remains approximately constant as a function of redshift between redshifts 1 and 2. Since the halo mass is evolving, such that a halo at $z=1.9$ will have a higher mass at $z=1.1$, we must be seeing growth in the central galaxy mass at close to the same rate as the halo mass is growing. \citet{Brown08} see a similar very slow evolution in the halo mass stellar mass relationship for red galaxies at $z < 1$ in the NDWFS. \citet{Zheng05} do observe a small amount of evolution in the halo mass luminosity relationship for central galaxies at $z < 1$, using samples from the SDSS and DEEP2, although the use of luminosities measured in different rest frame bands complicates the interpretation. 

\citet{Zehavi10} propose a parameterization of the stellar mass-to-halo mass relation for central galaxies consisting of a power-law component at high halo masses and a exponentially declining component at low halo masses given by 

\begin{equation}
\label{eq:SMHM}
	M_{*cen} = A\left(\frac{M_h}{M_t}\right)^{\alpha_M}exp\left(-\frac{M_t}{M_h} + 1\right)
\end{equation}
\\
where $A$, $M_t$ and $\alpha_M$ are free parameters.  $\alpha_M$ is the power-law slope, $M_t$ is the transition halo mass marking the point where the dominance of the two components switch, and $A$ is the normalization giving the median stellar mass at $M_t$. This form is well fit to the HOD measurements at $z\sim0.1$ where there is a clear shoulder in the stellar mass-to-halo mass relationship at around $10^{12}$ \hMsun \citep{Zehavi10}.

We fit this relation to the data shown in Figure \ref{fig:SMHM} and find best fit values of $A = 2.4_{-1.5}^{+0.7} \times 10^{10} \Msun$, $\alpha_M = 0.74_{-0.20}^{+0.38}$ and $M_t = 172_{-70}^{+45} \times 10^{10} \Msun$ shown as the solid line. Due to the limited range in stellar mass that our samples cover this fit is quite poorly constrained, although we might expect to be covering the transition region. We find that the best fit transition mass at $1 < z < 2$ is a factor of 4 higher than that found at $z \sim 0.1$ in the SDSS  \citep{Zehavi10}.  

\subsection{The Halo Occupation of Satellite Galaxies}

\begin{figure}
\vspace{12.0cm}
\includegraphics{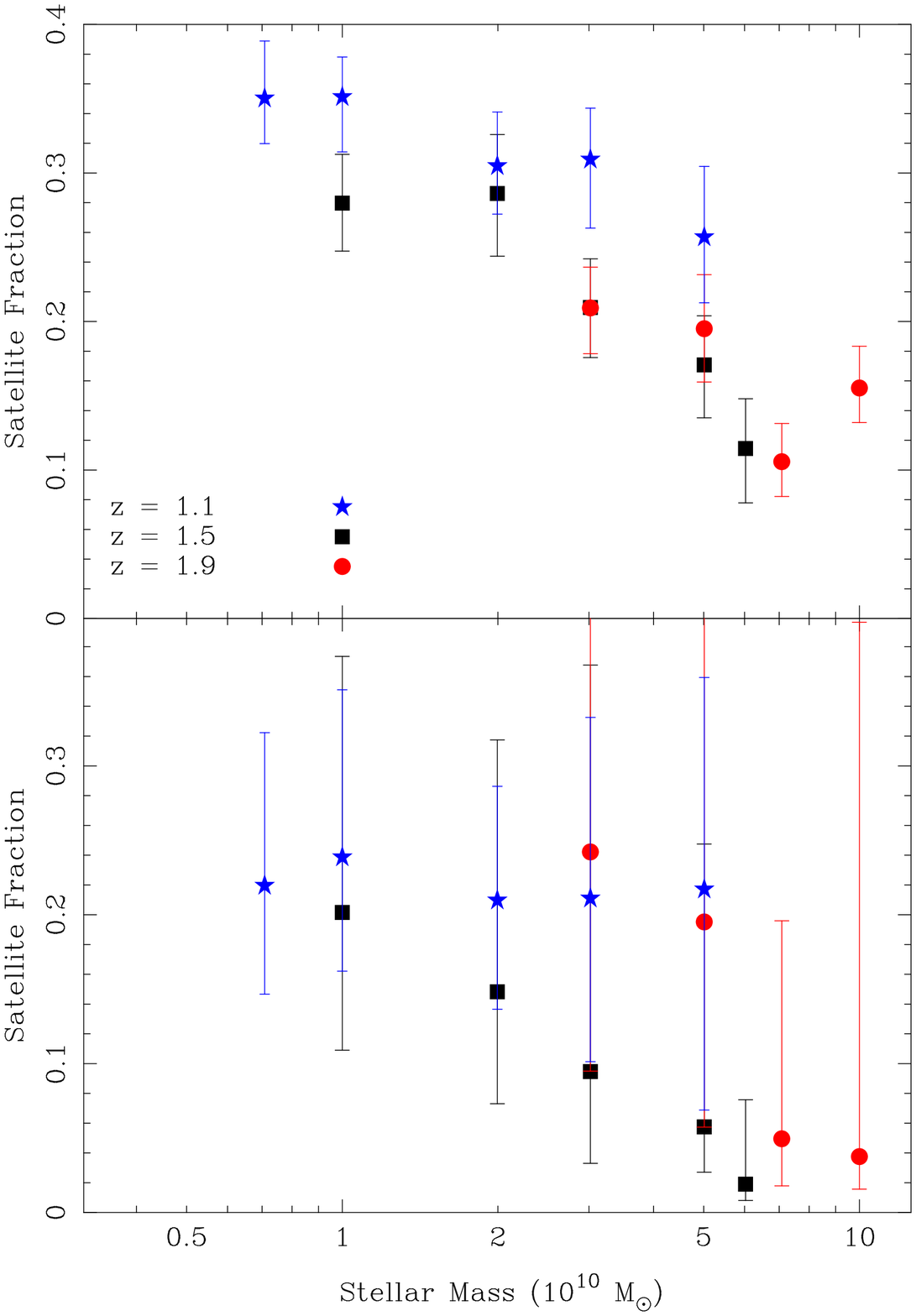}
\caption{\small The relationship between the fraction of galaxies above a stellar mass limit that are satellites in halos (rather than centrals) and the stellar mass limit determined from the HOD fits to the clustering (bottom) and the clustering and density (top).  In both panels the satellite fraction appears to reduce as the stellar mass increases. Including the density in the fit results in higher satellite fractions.  
\label{fig:Fsat}}
\end{figure}

Figure \ref{fig:SMHM} shows both the central mass scale $\Mmin$ and the satellite mass scale $M_1$ as a function of stellar mass, where $\Mmin$ is the halo mass at which a halo hosts on average 0.5 central galaxies and $M_1$ is the halo mass hosting on average 1 satellite galaxy. The dashed line in Figure \ref{fig:SMHM} shows the best fit relation to $\Mmin$ scaled by a factor of 6, which appears to fit the $M_1$ stellar mass relation pretty well. This offset implies that a halo hosting two galaxies, one central and one satellite, is typically more than 6 times as massive as a halo hosting just one central galaxy. Halos with masses in between $\Mmin$ and $M_1$ will typically host a single central galaxy with an above average stellar mass.

Observations from the SDSS at $z \sim 0.1$ find $M_1 \simeq 17\Mmin$ \citep{Zheng07,Zehavi10}, almost three times the difference we observe at $z > 1$. There are two reasons for this: First, our samples have higher stellar mass limits than the typical samples from the SDSS, and second, we expect there to be evolution with redshift. The stellar mass limit is important as the $M_1/\Mmin$ ratio is seen to become rapidly smaller for higher stellar mass (luminosity) limits \citep[e.g. see Figure 12 of][]{Zehavi10}. This dependence on halo mass is thought to be caused by the fact that the more massive the halo the later it is expected to form at any given epoch, so that there is less time for satellites to merge on to the central galaxy in more massive halos.

Even though the samples in \citet{Zehavi10} are luminosity limited, rather than mass limited, and at a different redshift we can make a reasonably good comparison by matching the space densities. In doing this we are assuming that the N most massive galaxies at one redshift correspond to the N most massive at another. At over-lapping space densities the $M_1/\Mmin$ ratio varies from 17 to 6, with a typical value of $\simeq 10$ at $z \sim 0.1$, and is much more consistent with our values for the lowest space densities (highest masses). We don't see much of a dependence of this ratio on stellar mass within our data, which could simply be the result of the relatively small stellar mass range of our samples or it could show a genuine reduction in the stellar mass dependence at high redshift. 

The expected redshift dependence has a similar origin as the stellar mass dependence, in that for any given halo mass the time with which satellites have to merge with the central galaxy is less at earlier epochs. \citet{Kravtsov04} use high resolution dissapationless N-body simulations to investigate the HOD and predict that $M_1/\Mmin$ should have 2/3 of its $z = 0$ value by $z = 1$ and 1/3 by $z = 3$. This prediction is roughly consistent with the reduction we observe compared to $z\sim0.1$.

Figure \ref{fig:Fsat} shows the relationship between satellite fraction and stellar mass limit for the HOD fits to the clustering ({\it bottom}) and the clustering and density combined ({\it top}). The HOD fit that includes the density shows higher satellite fractions than the fits to the clustering alone, although for most samples the significance is marginal. The reason for this difference is simple, for a given space density HODs with high satellite fractions are preferentially placing galaxies as satellites in high mass halos, which are more clustered, rather than centrals in lower mass halos, which are less clustered. For our samples, this allows the clustering to remain high, even though the space density is large. Of course there is a limit to how many satellite galaxies are allowed, determined by the amplitude of the small scale clustering.

Both fits show a general trend of decreasing satellite fraction with increasing stellar mass limit, a trend which has previously been seen in clustering measurements, group catalogs and from lensing at $z < 1$ \citep[e.g.][]{Zehavi05,Zheng07,Mandelbaum06,Yang08}. This trend simply results from the shape of the halo mass function; at high halo masses, the halo mass function has an exponential decline, so for a fixed ratio of $\Mmin$-to-$M_1$ halos with mass $M_1$ are increasingly rare relative to halos of mass $\Mmin$ as these mass scales increase. At lower halo masses the mass function flattens and the rate at which the satellite fraction changes decreases as the stellar mass limit decreases. This flattening in the satellite fraction is observed at $z \sim 0.1$ in the SDSS \citep{Zheng07,Yang08,Zehavi10} and there is evidence of this flattening in the measurements presented here for stellar mass limits less than $2\times10^{10}\Msun$.

\subsection{Star formation efficiency in central galaxies}

\begin{figure}
\vspace{12.8cm}
\includegraphics{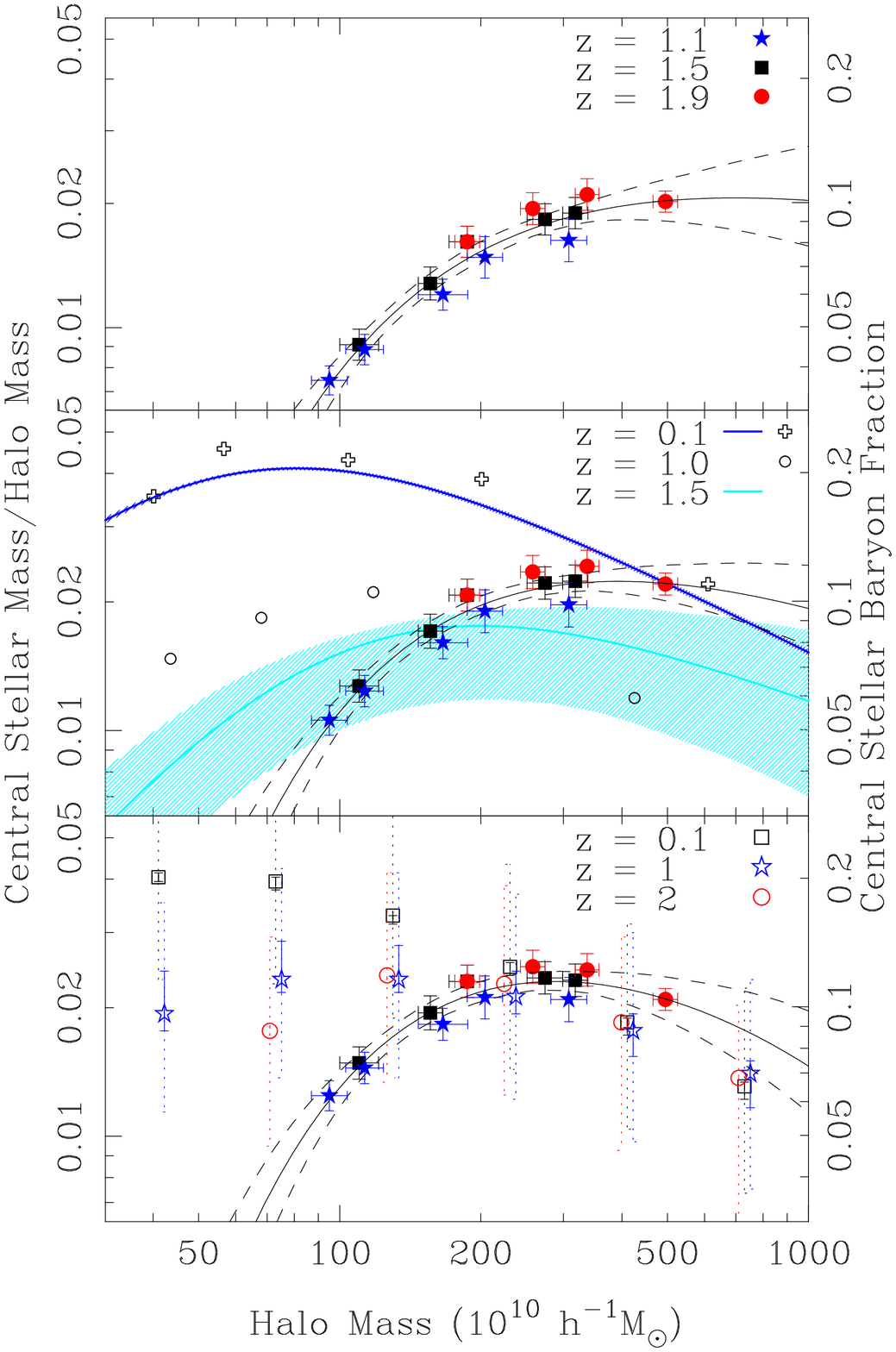}
\caption{\small Top panel: The relationship between the ratio of stellar mass-to-halo mass and halo mass for central galaxies determined from the HOD fits to the clustering and density. This ratio is proportional to the efficiency with which baryons have been converted to stars. The solid line shows the best fit relation of the form given in equation \ref{eq:SMHMrat} determined from the fit to the stellar mass-to-$\Mmin$ relation shown in Figure \ref{fig:HODSM}. The dashed lines show the $1\sigma$ confidence region on this fit. The star formation efficiency increases as the halo mass increases, before plateauing at high halo mass.
 Middle and bottom panels: Comparisons to similar measurements in the literature at $z=0.1, 1, 1.5$ and $2$ from \citet{Zheng07}, \citet{Moster10} (middle panel) and \citet{Behroozi10} (bottom panel). We have adjusted our measurements in these panels to match the stellar population modeling assumptions used in the literature determinations (see text for details). 
The peak in the relation, which corresponds to the peak in efficiency in the conversion of baryons to stars in the central galaxy, has shifted to higher halo masses at higher redshift.
\label{fig:SMHMrat}}
\end{figure}

A more physically intuitive way to view the stellar mass-to-halo mass relation for central galaxies is to consider how the ratio of the central galaxy stellar mass-to-halo mass depends on halo mass. Assuming that the baryon fraction is constant with halo mass then this ratio gives the efficiency with which baryons have been converted to stellar mass in the central galaxy. The top panel of Figure \ref{fig:SMHMrat} shows this ratio as a function of halo mass for our three samples along with the analytic relation derived from equation \ref{eq:SMHM} (Zehavi et al. 2010) 

\begin{equation}
\label{eq:SMHMrat}
	\frac{M_h}{M_{*cen}} =  \frac{M_t}{A}\left(\frac{M_h}{M_t}\right)^{1-\alpha_M}exp\left(\frac{M_t}{M_h} - 1\right)
\end{equation}

with $A$, $M_t$ and $\alpha_M$ the best fit parameters from before. Here $A/M_t$ is the halo mass-to-stellar mass ratio at $M_t$. This analytic relation is shown as the solid line in Figure \ref{fig:SMHMrat} with the 1$\sigma$ confidence region shown by the dashed lines. We see a steady rise in the star formation efficiency as halo mass increases, which plateaus at high halo masses and shows a hint of a turning over in the highest mass bin. We see little evidence of redshift evolution within our sample, although the turnover may be occurring slightly early at lower redshift.

In the middle and bottom panels of Figure \ref{fig:SMHMrat} we compare to other measurements from the literature at $z=0.1$, $z=1$, $z=1.5$ and $z=2$. The open crosses and circles in the middle panel are from \citet{Zheng07} and are based on HOD fits to clustering measurements in the SDSS and DEEP2 surveys. The blue and cyan shaded area in the middle panel show the $1\sigma$ confidence interval of this relation derived by sub-halo abundance matching (SHAM) to the SDSS stellar mass function at $z=0.1$ and stellar mass function measurements from \citet{Fontana06} in the Chandra Deep Field South at $z=1.5$ from \citet{Moster10}.

The bottom panel again shows further SHAM measurements from \citet{Behroozi10} who again make use of the SDSS stellar mass function at $z=0.1$, but use more recent stellar mass functions from \citet{Perez-Gonzalez08} and \citet{Marchesini09} at higher redshift. The solid errors on these points show the random errors on the measurements with the dotted errors showing the systematic errors.

All of the literature measurements use \citet{Bruzual03} stellar population synthesis models rather than the \citet{Maraston05} models that we have used. We have thus applied a correction to our stellar masses in the middle and bottom panels to account for this difference. \citet{Behroozi10} also assume a dust model from \citet{Blanton07} rather than \citet{Calzetti00} and so we further adjust of results in the bottom panel to reflect this. 

 All three SDSS relations show a characteristic maximum star formation efficiency at a halo mass of around $6\times10^{11}$ \hMsun with a steep decline either side, corresponding to a fall in efficiency at higher or lower halo mass. This peak efficiency has also been seen in several other studies at low redshift using the HOD, the Conditional Luminosity Function, group catalogs, SHAM and weak lensing \citep{Yang03,Eke05,Tinker05,Vale06,Mandelbaum06,Guo10}. 

When compared with these $z\simeq0$ results our measurements show that the halo mass-to-stellar mass vs. halo mass relation has changed significantly with redshift, with the peak in the star formation efficiency moving to higher halo masses at higher redshift. It is unfortunate that the area of our survey does not permit us to probe to slightly higher stellar masses, and thus halo masses, as it is not clear that we are definitely seeing the expected down turn in $M_*/M_h$ at high halo masses. However, there is a definite flattening in all three redshift ranges and some evidence of a down turn in the highest mass bin in the $\bar{z}$ = 1.9 sample. This turn over becomes more apparent in the bottom panel after the corrections to stellar population model and dust law are applied.

The literature measurements also show a shift in the peak efficiency to higher halo masses at higher redshift. These are broadly consistent with our measurements, although our peak efficiency is at somewhat higher halo masses and the rate of down turn at low halo masses appears to be faster. 

It is not surprising that there is not an exact agreement with \citet{Zheng07} since they use luminosity limited samples, limited in the rest B band at $z\sim~1$, resulting in a color dependent selection that is not stellar mass limited. Cosmic variance caused by the small field used for the stellar mass function measurement in \citet{Moster10}, along with small differences in the stellar mass determinations and cosmology could easily explain the difference between our relations.

Our measurement at $z=1$ lies just about within the systematic error of \citet{Behroozi10}, however we have adjusted our stellar masses to match their model, and uncertainties in the stellar population synthesis model masses are the dominant factor in their systematic error determination. We assume the same cosmology and whilst there are slight differences in our halo mass definitions one would expect better agreement, particularly as the NMBS data produce stellar mass functions which very closely match those used by \citet{Behroozi10}. The largest difference is for lowest stellar mass limited samples where our best fit halo masses would need to be about 30\% lower than \citet{Behroozi10}. The cause of this difference appears to be the fact that we are fitting the clustering as well as the space density of galaxies, whereas the SHAM just uses the space density of galaxies. For a given stellar mass limit we could easily reduce the value of $\Mmin$ in the halo model (which gives the median central stellar mass) if we made $\M1$ higher thus reducing the satellite fraction. However, this would reduce the clustering amplitude on both small and large scales and be incompatible with our clustering measurements. At lower stellar masses the SHAM model is assigning fewer galaxies to be satellites and more to be centrals than appears to be compatible with our clustering measurements.  This illustrates the importance of testing the assumptions that go into the SHAM model with additional observations such as the clustering.

Returning to the redshift evolution of the star formation efficiency the shift in the overall normalization of this relation just reflects the continued conversion of baryons into stars. If we assume a universal baryon fraction of 16.9\% then we can convert the halo mass-to-stellar mass ratio to the fraction of baryons converted to stars in the central galaxy. At $z\sim1.5$ this stellar baryonic fraction peaks at about 10\%, whereas at $z=0$ it peaks at 25\%. At high halo masses the high redshift and low redshift relations approach each other. This means that central galaxies in high mass halos are growing much more slowly in stellar mass than their host halos are growing in dark matter mass. It does not mean that the total fraction of baryons in stars has hardly changed, just that these stars are not being effectively added to the central galaxy and may be in either satellites or in the diffuse intra halo background. The opposite seems to be the case at lower halo masses where there appears to be an increasing rate of stellar mass build up for centrals, compared to the rate of growth of their parent halos. This is in a sense another manifestation of the `downsizing' paradigm, but for halos. The most massive halos are more efficient in converting baryons to stars at high redshift and are becoming increasingly inefficient in doing so as time goes on compared to lower mass halos.\\

\section{Summary and Conclusions}
\label{sec:conclusions}

We present here a detailed analysis of the clustering of galaxies as a function of their stellar mass at $1 < z < 2$ using data from the NEWFIRM Medium Band Survey. The precise nature of the NMBS photometric redshifts allows us to define samples that are very close to being volume limited with accurate stellar mass estimates. We find the following:\\

(i) In all three redshift slices we see a significant dependence of the clustering amplitude on the stellar mass limit of the sample, both when comparing the 2-point angular correlation functions directly, or when considering the amplitude of a power law fit to the correlation function measurements. This shows that the strong stellar mass dependent clustering seen at $z\lesssim1$ persists up to $z\sim2$.

(ii) We fit halo models to our measurements using the form of the HOD from \citet{Zheng05}, fitting for the central and satellite mass thresholds $\Mmin$ and $\M1$ respectively. We find that both $\Mmin$ and $\M1$ show a significant increase with increasing stellar mass limit, confirming that at strong stellar mass-to-halo mass relationship is in place at $1 < z < 2$.

(iii) For the HOD definition that we use, the stellar mass limit of our samples corresponds to the median stellar mass of central galaxies hosted by halos of mass $\Mmin$. We see little evidence for any evolution in this relationship between central stellar mass and halo mass within our sampled redshift range, although there is clear evidence that it evolves from the redshift zero relation.

(iv) We determine the efficiency with which baryons are converted into stars in central galaxies as a function of halo mass by calculating the stellar mass-to-halo mass ratio. We see a peak star formation efficiency in halos of mass $\sim 3\times10^{12}h^{-1}\Msun$ with a clear decrease in efficiency at low halo masses. There is some weak evidence of a downturn at high halo masses, but this conclusion is limited since the size of our survey prevents us from probing stellar mass limits greater than $10^{11}\Msun$ due to the small number of galaxies with these high masses. Measurements at $z \simeq 0$ show a peak efficiency in halos of $\sim 7\times10^{11}h^{-1}\Msun$, providing clear evidence of a shift in peak efficiency to higher halo masses at higher redshift. This halo `downsizing' is a similar phenomenon as the galaxy `downsizing' seen as a function of galaxy mass.

(v) We find evidence that the fraction of satellite galaxies increases as the stellar mass limit decreases in a similar manner to that seen at $z\simeq0$. The ratio between the central and satellite mass thresholds $\Mmin$ and $\M1$ remains approximately constant over the relatively narrow range in stellar mass that we probe. The ratio does appear to have decreased compared with the $z\simeq0$ values, consistent with the basic expectations of N-body simulations.  

(vi) We show some evidence that the halo model is unable to reconcile both the observed clustering and space density of highly biased galaxies at $z > 1$. The significance of this is marginal, and it is possible that we could have been unlucky with cosmic variance; but taken with other similar recent observations \citep{Matsuoka10} it may imply that there is an incompleteness in the halo model, for instance the halo bias relation may need to be adjusted in the high bias regime or galaxies may not be distributed within halos following an NFW profile.
 
There remains much scope for furthering this work in the near future.
The forthcoming NEWFIRM Medium Band Survey II, which will cover 10 times the area as NMBS in 5 fields, but to shallower flux limits, will enable us to extend this work to higher stellar masses. This should allow the down turn in the stellar mass-to-halo mass ratio halo mass relation to be precisely pin-pointed as well as resolve whether there is an issue with the halo model at high redshift and high bias.

The ultimate goal of this work will be to use the halo model framework to place constraints on the evolution of the galaxy population, by combining these measurements with those at lower redshifts as has been done by e.g. \citet{Zheng07,White07,Wake08a,Brown08}. We have already begun this process and in a forthcoming paper we will combine the measurements presented here at $1 < z < 2$ with measurements of clustering as a function of stellar mass in the SDSS at z = 0.1, where we have defined the stellar mass in a consistent manner (Wake et al. 2010 in prep.). This will allow us to see if the stellar mass halo mass relation remains constant over this range and, by combining it's evolution with the evolution of the halo mass, determine how central galaxies have grown as a function of cosmic time over the last 10Gyrs.  

\acknowledgments{

DAW would like to thank Ravi Sheth, Zheng Zheng, Micheal Brown, Carlton Baugh, Richard Bower, Scott Croom, Darren Croton and Frank van den Bosch for helpful discussions on this work. Ron Probst and the NEWFIRM team are thanked
for their work on the instrument and help during the observations.
We used the HyperLeda database (http://leda.univ-lyon1.fr).
This paper is partly
based on observations obtained with MegaPrime/MegaCam, a joint project
of CFHT and CEA/DAPNIA, at the Canada-France-Hawaii Telescope (CFHT)
which is operated by the National Research Council (NRC) of Canada,
the Institut National des Science de l'Univers of the Centre National
de la Recherche Scientifique (CNRS) of France, and the University of
Hawaii. We thank the CARS team for providing us with their
reduced CFHT mosaics. Support from NSF grants AST- 0449678 and AST-0807974 is gratefully acknowledged. 
}

\appendix

\section{Estimating errors on the 2-point correlation function}
\label{sec:sims}

\begin{figure*}
\vspace{8.0cm}
\includegraphics{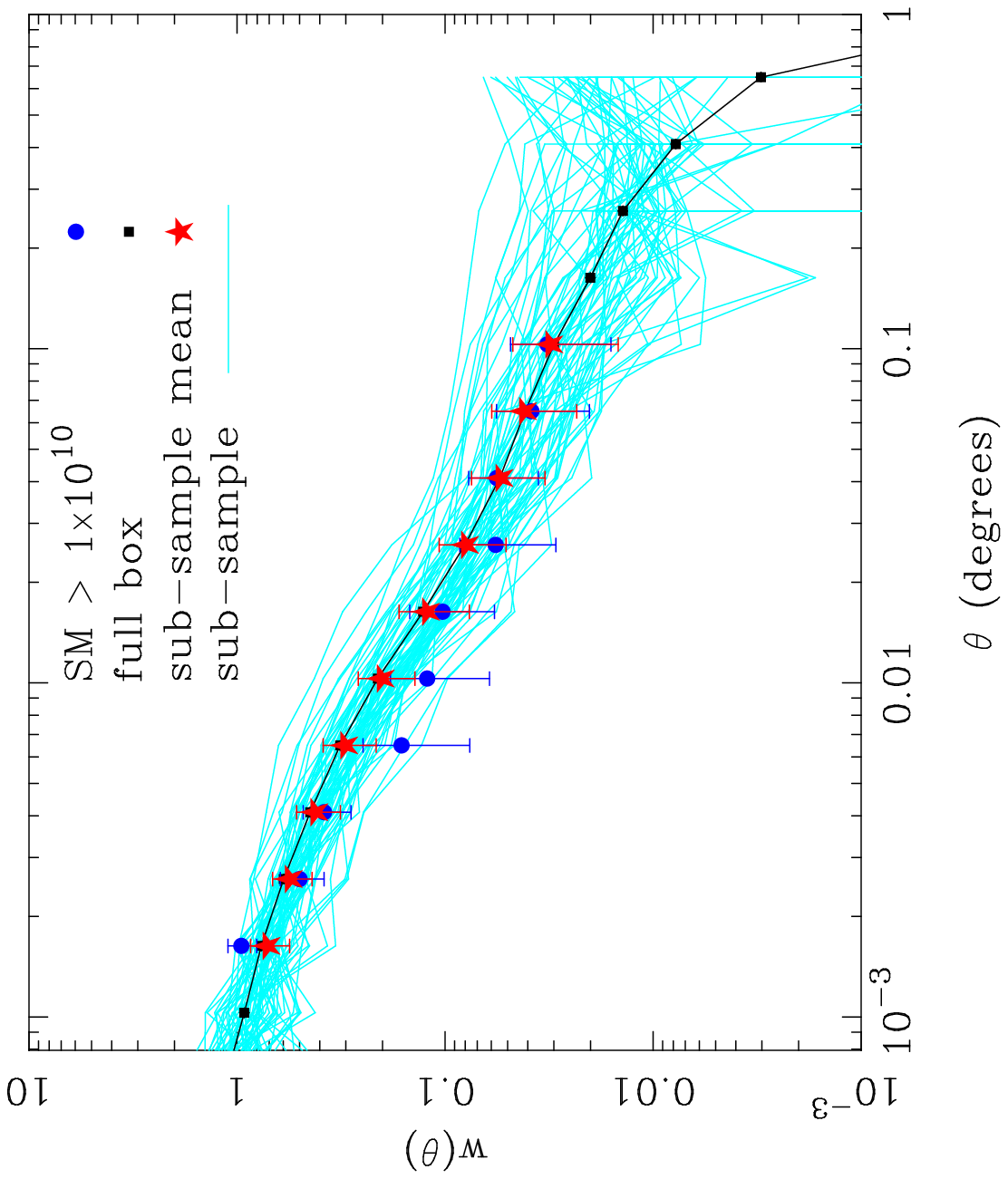}
\includegraphics{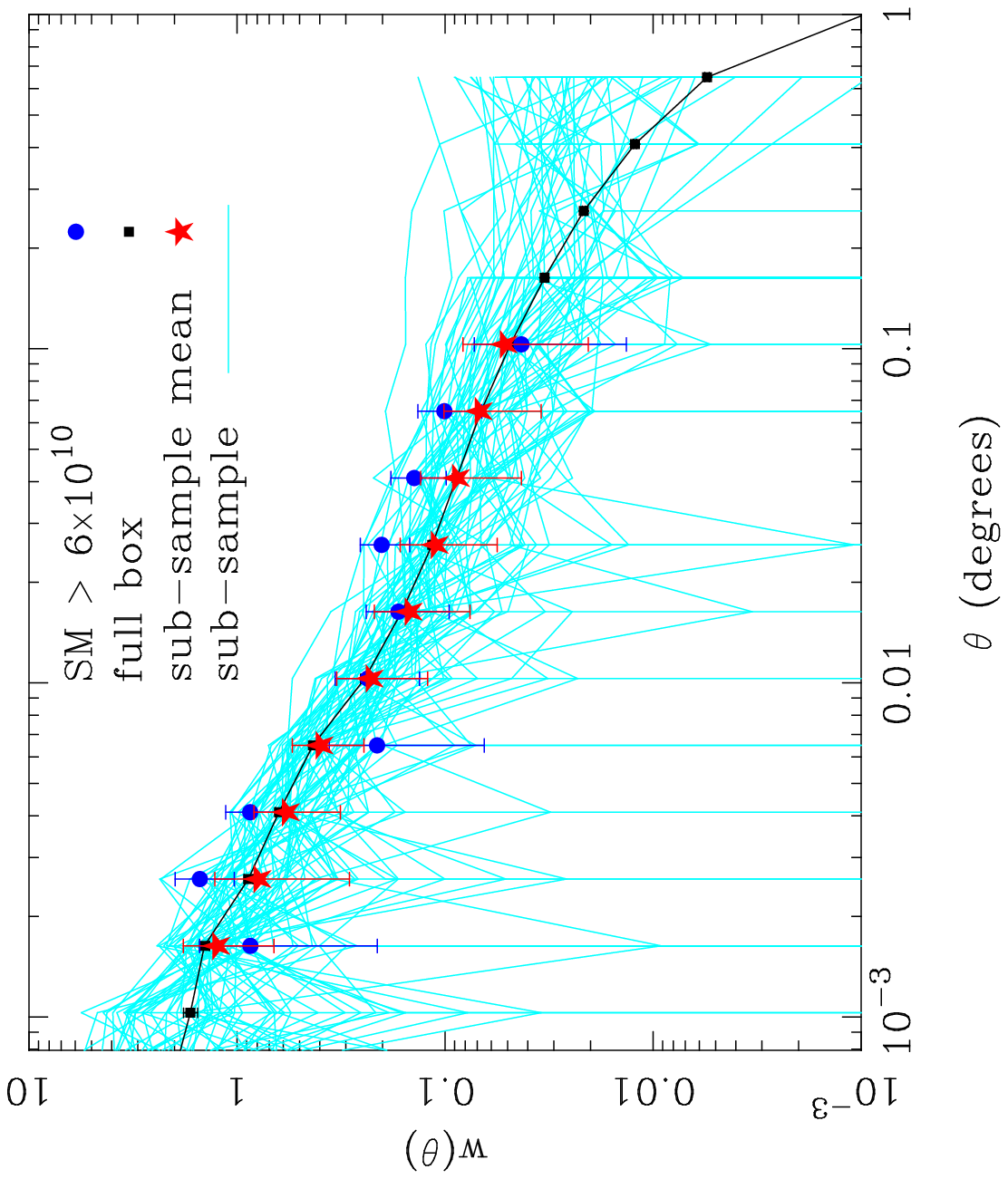}
\caption{\small The measured and mock angular correlation functions for the most and least massive galaxies in the $\bar{z} = 1.5$ sample. The blue points show the measurements from the NMBS, the black squares the full box mock, the red stars the mean of the individual survey mocks, and the cyan lines the individual survey mocks. All have been corrected for the integral constraint where appropriate.
\label{fig:millfit}}
\end{figure*}

Making an accurate estimation of the errors on a correlation function measurement is particularly challenging, and there are several established methods in the literature, all of which have advantages and disadvantages. Since the individual data points in any correlation function are highly correlated, particularly on large scales, it is vital to estimate the full covariance when determining the errors. This immediately rules out the use of Poisson errors, which despite these problems are still often used. 

This leaves two broad classes of method: internal estimators such as bootstrap or jackknife resampling techniques that make use of the data themselves, and mock catalogs, whereby multiple fake data sets are created and compared. The advantage of the internal estimators, is that they include all of the real correlations in the data, something which mock catalogs may not. The disadvantage is that they are of limited size and so can produce noisy estimates of the covariance matrix and may also show dependence on the numbers and size of the sub-regions used when re-sampling \citep{Norberg09}.

We made an initial attempt to use the jackknife resampling technique to estimate the covariance on our measurements. We define multiple sub-regions with equal area and then recalculate the correlation function removing one subregion at a time to calculate the covariance. We perform this procedure multiple times splitting our data in to 9, 25, 49 and 100 sub-regions. Comparing the covariances we find a strong dependence of the magnitude of the errors on the number of sub-regions we use in the jackknife calculation, particularly on large scales, an effect reported previously in Norberg et al. (2009; but see also Zehavi et al. 2005). The strong large scale dependence is not surprising since the size of our sub-regions is comparable or smaller than the scale at which we wish to estimate the error. However, we still see a systematic correlation at smaller scales, such that the errors increase as the number of sub-samples increases. For this reason we believe that mock catalogs provide a significantly better approach for our analysis.

We choose to use the Millennium simulation \citep{Springel05} to generate our mock surveys, which is almost ideal for our purposes. For each redshift sample we download a halo catalog with halo masses $> 7\times10^{10}$\hMsun~ at the output with the redshift closest to the mean of our sample, redshifts 1.08, 1.50 and 1.91 respectively. 

We then make a HOD fit to the correlation function and density of each of our samples using the method described in Section \ref{sec:halo}. We make use of the covariance matrices generated using the jackknife method with 9 sub-samples. There is considerable noise in these covariance matrices due to the small number of jackknifes and we mitigate this by restricting the fit to use just 95\% of the covariance based on a principle component analysis following the techniques proposed by \citet{Porciani06,Norberg09}. This is more than adequate to generate the mocks, where we are just interested in the best fitting HOD and not concerned with confidences etc. 

We fit to both the density and clustering as it is important that we get the best match to both in the mocks. If we just fit to the clustering, something we tested, we see that the density in the mocks is much lower than in our samples, particularly at high stellar masses and at $\bar{z}=1.5$. This leads to a substantial over estimation of the errors simply as a result of increased shot noise. 

The Millennium simulation has $\sigma_8$ = 0.9, where as the currently favored value, and the one we assume, is 0.8. When fitting the halo model in this instance we therefore use $\sigma_8$ = 0.9. Despite this difference the resulting mocks are able to reproduce the clustering and space density of our samples with the same accuracy as the $sigma_8$ = 0.8 HOD fits and so we are happy to use these to estimate our errors.

The dark matter halo catalogs extracted from the Millennium simulation are then populated following the best fitting HOD for each sample, placing the central galaxies at the center of the halos and distributing the satellites following an NFW profile. The whole Millennium box is then split up into multiple copies of our full survey geometry. The Millennium box size of 500$h^{-1}Mpc$ is not quite large enough to accommodate the full size of our samples in the redshift direction and so we wrap the box in order to accomplish this. We are able to generate 170, 128, and 98 realizations of a single field at redshifts 1.1, 1.5 and 1.9 respectively. When generating the mocks for both fields combined we ensure that each pair of fields is separated by at least 100$h^{-1}Mpc$ so they are not correlated.

The angular correlation function is then calculated for the full box mock, and in each of the survey mocks. The covariance for each stellar mass limited sample
is determined from the multiple survey mocks. We make a direct estimate of the integral constraint for the mock surveys by comparing the mean correlation function of the survey mocks to the correlation function of the full box mock.

Figure \ref{fig:millfit} shows the results of this process for the lowest and highest stellar mass limited samples at $z = 1.5$ with both fields combined. The blue points on each plot show the correlation function measurement for the NMBS sample, the black squares show the correlation function from the full box mock, and the red stars indicate the mean of the individual survey mocks corrected by the integral constraint. The cyan lines show the individual correlation function measurements from each survey mock. In all cases the mean survey mocks agree with the full box mock as expected. As discussed in Section \ref{sec:tension} for the high stellar mass sample the clustering measurement is higher than that determined from the mock, although there are some individual survey mocks that do show this level of clustering.

The errors estimated in this way treat each stellar mass limited sample as if it were independent of the others in our survey. As such we estimate the error one would expect if the correlation function was determined at a given stellar mass limit in a given redshift interval on any piece of sky. When comparing our stellar mass limited samples with each other they are not independent as they are measured in the same volume each time (the same piece of sky). It is almost certainly the case that if in one particular volume the clustering amplitude of galaxies more massive than $1\times10^{10}\Msun$ is higher than average the clustering amplitude of galaxies more massive than $5\times10^{10}\Msun$ will also be higher. Our error estimates ignore this and we treat our measurements as if they had been made in independent volumes when testing for stellar mass dependent clustering. In this case we are most likely overestimating the error on any mass dependent clustering and underestimating its significance. As such we have estimated the error in a most conservative fashion and we still find significant stellar mass dependent clustering. We therefore choose not to pursue the much more complex simulations that would be required to fully model this effect. Of course our error estimates are perfectly correct if one wished to compare our measurements with similar ones estimated from another region of sky.

\section{Halo model}
\label{sec:AppHalo}

Here we present a more detailed overview of our halo model calculation.\\

The fraction of halos of mass $M$ which host centrals is modeled as 
\begin{equation}
\label{eq:AppNcen}
	\langle N_c|M\rangle = \frac{1}{2}\left[1 + \erf\left(\frac{logM - logM_{min}}{\sigma_{logM}}\right)\right] 
\end{equation}

where erf is the error function $M$ is the halo mass and $\Mmin$ and $\sigM$ parameterize the HOD.
Only halos which host a central may host satellites.  In such halos, the number of satellites is drawn from a Poisson distribution with mean 
\begin{equation}
 \label{eq:AppNsat}
	\langle N_s|M \rangle = \left(\frac{M - M_0}{M_1^\prime}\right)^{\alpha}.
\end{equation}
Thus, the mean number of galaxies in halos of mass $M$ is 
\begin{equation}
 \label{eq:AppNtot}
 \langle N|M\rangle = \langle N_c|M\rangle[1 + \langle N_s|M \rangle], 
\end{equation} 
and the predicted number density of galaxies is 
\begin{equation}
\label{eq:Appden}
	n_g =  \int dM\, n(M)\, \langle N|M\rangle,
\end{equation}
where $n(M)$ is the halo mass function, for which we use the latest parameterization given by \citet{Tinker10b}.

We further assume that the satellite galaxies in a halo trace an NFW profile \citep{Navarro96} around the halo center, and that the halos are biased tracers of the dark matter distribution.  The halo bias ($b(M)$) depends on halo mass in a way that can be estimated directly from the halo mass function \citep{Sheth99}, and we use the most up to date parameterization of \citet{Tinker10b}. With these assumptions the halo model for $\xi(r)$ is completely specified \citep[e.g.][]{Cooray02}.  We then calculate $w(\theta)$ from $\xi(r)$ using equation~(\ref{eq:limber}). 

In \citet{Wake08a,Wake08b} we used the linear theory power spectrum ($P_{Lin}(k)$) throughout the calculation, whereas we now use the non-linear power spectrum when calculating the 2-halo term. We also apply the scale dependent bias and halo exclusion

In addition to $\xi(r)$, we are interested in the satellite fraction, 
\begin{equation}
\label{eq:AppFsat}
 F_{sat} = \int dM\, n(M)\,\langle N_c|M\rangle\,\langle N_s|M\rangle/n_g.  
\end{equation} 
and two measures of the typical masses of galaxy host halos: 
an effective halo mass 
\begin{equation}
\label{eq:AppMeff}
	M_{eff} =  \int dM\, M\, n(M)\, \langle N|M\rangle/n_g,
\end{equation}
and the average linear bias factor 
\begin{equation}
\label{eq:Appblin}
	b_{g} =  \int dM\, n(M)\, b(M)\,\langle N|M\rangle/n_g,
\end{equation}
where $b(M)$ is the halo bias.

Our notation is intended to make explicit the fact that the mean number density of central-satellite pairs from such halos is $n(M)\,\langle N_c|M\rangle\,\langle N_s|M \rangle$, and the mean number density of distinct satellite-satellite pairs is $n(M)\,\langle N_c|M\rangle\,\langle N_s|M \rangle^2/2$ (because we are assuming the satellite counts are Poisson).  

Our model for the real-space 2-point function is 
\begin{equation}
 \xi(r) = 1+\xi_{cs}(r) + 1+\xi_{ss}(r) + \xi_{2h}(r)
\end{equation}
where 
\begin{equation}
1+\xi_{cs}(r) = \int dM\, {n(M)\langle N_c|M\rangle\over n_g}\, \langle N_s|M\rangle\, {\rho(r|M)\over n_gM} 
\end{equation}

\begin{equation}
1+\xi_{ss}(r) = \int dM\, {n(M)\langle N_c|M\rangle \over n_g}\, \frac{\langle N_s|M\rangle^2}{2}\, {\lambda(r|M)\over n_gM^2} 
\end{equation}
and 
\begin{equation}
\xi_{2h}(r) = \int \frac{dk}{k}\,\frac{k^3P_{2h}(k)}{2\pi^2}\frac{sinkr}{kr}
\end{equation}
with 
\begin{equation}
 P_{2h}(k,r) = b_g(k,r)^2\,P(k),
\end{equation}
where
\begin{equation}
\label{eq:bg}
 b_g(k,r) = \int_0^{M_{lim}(r)} dM {n(M)\over n_g^\prime}\, b(M,r)\,
 \langle N_c|M\rangle\Bigl[1 + \langle N_s|M\rangle u(k|M)\Bigr].
\end{equation}
In the expressions above,
 $\rho(r|M)$ is the density profile of halos of mass $M$,
 $\lambda(r|M)$ denotes the convolution of two such profiles, 
 $u(k|M)$ is the Fourier transform of $\rho(r|M)/M$, 
 and $P(k)$ denotes the non-linear theory power spectrum at the redshift of interest.  
The scale dependent bias ($b(M,r)$) is calculated following \citet{Tinker05} as 
\begin{equation}
	b^2(M,r) = b^2(M)\frac{[1+1.17\xi_m(r)]^{1.49}}{[1+0.69\xi_m(r)]^{2.09}}
\end{equation}
where $\xi_m(r)$ is the non-linear real-space matter 2pt correlation function. The upper limit to the integral, $M_{lim}(r)$, as well as the use of the {\it restricted} number density, $n_g^\prime$ in equation \ref{eq:bg}, take into account the effect of halo exclusion, with  $M_{lim}(r)$ determined following the method of \citet{Tinker05} and 
\begin{equation}
	n_g^\prime = \int_0^{M_{lim}(r)} dM\, n(M)\, \langle N|M\rangle.
\end{equation}		
 
All these quantities, along with the mass function $n(M)$ and bias 
factor $b(M,r)$, are to be evaluated at the redshift of interest.  
We have already specified how, for a given halo mass, the virial 
radius depends on redshift; the NFW halo density profile is also 
specified by its concentration, for which we assume the relation of 
\citet{Bullock01}.
All this, in the right hand side of equation~(\ref{eq:limber}), gives 
the halo model calculation of $w(\theta)$.

\clearpage


\begin{thebibliography}{}
\bibitem[Abazajian et al.(2009)]{Abazajian09} Abazajian, K.~N., et al.\ 2009, \apjs, 182, 543
\bibitem[Abbas et al.(2010)]{Abbas10} Abbas, U., et al.\ 2010, \mnras, 406, 1306
\bibitem[Adelberger et al.(2005a)]{Adelberger05a} Adelberger, K.~L., 
Steidel, C.~C., Pettini, M., Shapley, A.~E., Reddy, N.~A., 
\& Erb, D.~K.\ 2005, \apj, 619, 697
\bibitem[Adelberger et al.(2005b)]{Adelberger05b} Adelberger, K.~L., 
Erb, D.~K., Steidel, C.~C., Reddy, N.~A., Pettini, M., 
\& Shapley, A.~E.\ 2005, \apjl, 620, L75
\bibitem[Auger et al.(2010)]{Auger10} Auger, M.~W., Treu, T., 
Bolton, A.~S., Gavazzi, R., Koopmans, L.~V.~E., Marshall, P.~J., Moustakas, 
L.~A., \& Burles, S.\ 2010, arXiv:1007.2880
\bibitem[Barmby et al.(2006)]{Barmby06} Barmby, P., et al.\ 2006, \apj, 642, 126
\bibitem[Baugh(2006)]{Baugh06} Baugh, C.~M.\ 2006, Reports of Progress in Physics, 69, 3101
\bibitem[Behroozi et al.(2010)]{Behroozi10} Behroozi, P.~S., Conroy, C., \& Wechsler, R.~H.\ 2010, \apj, 717, 379
\bibitem[Berlind \& Weinberg(2002)]{Berlind02} Berlind, A.~A., \& Weinberg, D.~H.\ 2002, \apj, 575, 587 
\bibitem[Bertin \& Arnouts(1996)]{Bertin96} Bertin, E., \& Arnouts, S.\ 1996, \aaps, 117, 393
\bibitem[Bielby et al.(2010)]{Bielby10} Bielby, R., et al.\ 2010, arXiv:1005.3028
\bibitem[Blake et al.(2008)]{Blake08} Blake, C., Collister, A., 
\& Lahav, O.\ 2008, \mnras, 385, 1257
\bibitem[Blanc et al.(2008)]{Blanc08} Blanc, G.~A., et al.\ 2008, \apj, 681, 1099
\bibitem[Blanton \& Roweis(2007)]{Blanton07} Blanton, M.~R., \& Roweis, S.\ 2007, \aj, 133, 734
\bibitem[Blumenthal et al.(1984)]{Blumenthal84} Blumenthal, G.~R., Faber, S.~M., Primack, J.~R., \& Rees, M.~J.\ 1984, \nat, 311, 517
\bibitem[Brammer et al.(2008)]{Brammer08} Brammer, G.~B., van Dokkum, P.~G., \& Coppi, P.\ 2008, \apj, 686, 1503
\bibitem[Brammer et al.(2009)]{Brammer09} Brammer, G.~B., et al.\ 2009, \apjl, 706, L173
\bibitem[Brown et al.(2005)]{Brown05} Brown, M.~J.~I., Jannuzi, 
B.~T., Dey, A., \& Tiede, G.~P.\ 2005, \apj, 621, 41
\bibitem[Brown et al.(2008)]{Brown08} Brown, M.~J.~I., et al.\ 2008, \apj, 682, 937
\bibitem[Bruzual \& Charlot(2003)]{Bruzual03} Bruzual, G., \& Charlot, S.\ 2003, \mnras, 344, 1000 
\bibitem[Budav{\'a}ri et al.(2003)]{Budavari03} Budav{\'a}ri, T., et al.\ 2003, \apj, 595, 59
\bibitem[Bullock et al.(2001)]{Bullock01} Bullock, J.~S., Kolatt, T.~S., Sigad, Y., Somerville, R.~S., Kravtsov, A.~V., Klypin, A.~A., Primack, J.~R., \& Dekel, A.\ 2001, \mnras, 321, 559
\bibitem[Calzetti et al.(2000)]{Calzetti00} Calzetti, D., Armus, L., Bohlin, R.~C., Kinney, A.~L., Koornneef, J., \& Storchi-Bergmann, T.\ 2000, \apj, 533, 682
\bibitem[Coil et al.(2004)]{Coil04} Coil, A.~L., et al.\ 2004, 
\apj, 609, 525
\bibitem[Coil et al.(2006)]{Coil06} Coil, A.~L., Newman, 
J.~A., Cooper, M.~C., Davis, M., Faber, S.~M., Koo, D.~C., 
\& Willmer, C.~N.~A.\ 2006, \apj, 644, 671
\bibitem[Coil et al.(2008)]{Coil08} Coil, A.~L., et al.\ 2008, 
\apj, 672, 153
\bibitem[Colless et al.(2001)]{Colless01} Colless, M., et al.\ 2001, \mnras, 328, 1039
\bibitem[Conroy et al.(2006)]{Conroy06} Conroy, C., Wechsler, 
R.~H., \& Kravtsov, A.~V.\ 2006, \apj, 647, 201
\bibitem[Conroy et al.(2007)]{Conroy07} Conroy, C., et al.\ 2007, \apj, 654, 153
\bibitem[Conroy \& Wechsler(2009a)]{Conroy09} Conroy, C., \& Wechsler, R.~H.\ 2009, \apj, 696, 620
\bibitem[Conroy et al.(2009b)]{Conroy09b} Conroy, C., Gunn, J.~E., 
\& White, M.\ 2009, \apj, 699, 486
\bibitem[Cooray \& Sheth(2002)]{Cooray02} Cooray, A., \& Sheth, R.\ 2002, \physrep, 372, 1
\bibitem[Daddi et al.(2000)]{Daddi00} Daddi, E., Cimatti, A., Pozzetti, L., Hoekstra, H., R{\"o}ttgering, H.~J.~A., Renzini, A., Zamorani, G., \& Mannucci, F.\ 2000, \aap, 361, 535 
\bibitem[Davis et al.(2007)]{Davis07} Davis, M., et al.\ 2007, \apjl, 660, L1
\bibitem[Eisenstein et al.(2001)]{Eisenstein01} Eisenstein, D.~J., et al.\ 2001, \aj, 122, 2267
\bibitem[Eke et al.(2005)]{Eke05} Eke, V.~R., Baugh, C.~M., 
Cole, S., Frenk, C.~S., King, H.~M., \& Peacock, J.~A.\ 2005, \mnras, 362, 1233
\bibitem[Erben et al.(2009)]{Erben09} Erben, T., et al.\ 2009, \aap, 493, 1197
\bibitem[Fall \& Efstathiou(1980)]{Fall80} Fall, S.~M., \& Efstathiou, G.\ 1980, \mnras, 193, 189
\bibitem[Fontana et al.(2006)]{Fontana06} Fontana, A., et al.\ 2006, \aap, 459, 745
\bibitem[Foucaud et al.(2007)]{Foucaud07} Foucaud, S., et al.\ 
2007, \mnras, 376, L2
\bibitem[Foucaud et al.(2010)]{Foucaud10} Foucaud, S., Conselice, C.~J., Hartley, W.~G., Lane, K.~P., Bamford, S.~P., Almaini, O., \& Bundy, K.\ 2010, \mnras, 406, 147
\bibitem[Grazian et al.(2006)]{Grazian06} Grazian, A., et al.\ 2006, \aap, 453, 507
\bibitem[Groth \& Peebles(1977)]{Groth77} Groth, E.~J., \& Peebles, P.~J.~E.\ 1977, \apj, 217, 385
\bibitem[Guo et al.(2010)]{Guo10} Guo, Q., White, S., Li, C., \& Boylan-Kolchin, M.\ 2010, \mnras, 404, 1111
\bibitem[Jing et al.(1998)]{Jing98} Jing, Y.~P., Mo, H.~J., \& Boerner, G.\ 1998, \apj, 494, 1
\bibitem[Hamana et al.(2001)]{Hamana01} Hamana, T., Yoshida, N., Suto, Y., \& Evrard, A.~E.\ 2001, \apjl, 561, L143
\bibitem[Hartley et al.(2008)]{Hartley08} Hartley, W.~G., et al.\ 2008, \mnras, 391, 1301
\bibitem[Hayashi et al.(2007)]{Hayashi07} Hayashi, M., Shimasaku, K., Motohara, K., Yoshida, M., Okamura, S., \& Kashikawa, N.\ 2007, \apj, 660, 72
\bibitem[Heymans et al.(2006)]{Heymans06} Heymans, C., et al.\ 2006, \mnras, 371, L60
\bibitem[Hildebrandt et al.(2009)]{Hildebrandt09} Hildebrandt, H., Pielorz, J., Erben, T., van Waerbeke, L., Simon, P., \& Capak, P.\ 2009, \aap, 498, 725
\bibitem[Infante(1994)]{Infante94} Infante, L.\ 1994, \aap, 282, 353
\bibitem[Kim et al.(2010)]{Kim10} Kim, J.~-., Edge, A.~C., 
Wake, D.~A., \& Stott, J.~P.\ 2010, arXiv:1007.5154 
\bibitem[Kong et al.(2006)]{Kong06} Kong, X., et al.\ 2006, \apj, 638, 72
\bibitem[Kong et al.(2009)]{Kong09} Kong, X., Fang, G., 
Arimoto, N., \& Wang, M.\ 2009, \apj, 702, 1458
\bibitem[Kravtsov et al.(2004)]{Kravtsov04} Kravtsov, A.~V., Berlind, A.~A., Wechsler, R.~H., Klypin, A.~A., Gottl{\"o}ber, S., Allgood, B., \& Primack, J.~R.\ 2004, \apj, 609, 35
\bibitem[Kriek et al.(2008)]{Kriek08} Kriek, M., et al.\ 2008, \apj, 677, 219
\bibitem[Kriek et al.(2009)]{Kriek09} Kriek, M., van Dokkum, P.~G., Labb{\'e}, I., Franx, M., Illingworth, G.~D., Marchesini, D., \& Quadri, R.~F.\ 2009, \apj, 700, 221
\bibitem[Kroupa(2001)]{Kroupa01} Kroupa, P.\ 2001, \mnras, 322, 231
\bibitem[Labb{\'e} et al.(2003)]{Labbe03} Labb{\'e}, I., et al.\ 2003, \aj, 125, 1107
\bibitem[Landy \& Szalay(1993)]{Landy93} Landy, S.~D., \& Szalay, A.~S.\ 1993, \apj, 412, 64
\bibitem[Lee et al.(2006)]{Lee06} Lee, K.-S., Giavalisco, M., 
Gnedin, O.~Y., Somerville, R.~S., Ferguson, H.~C., Dickinson, M., 
\& Ouchi, M.\ 2006, \apj, 642, 63
\bibitem[Lee et al.(2009)]{Lee09} Lee, K.-S., Giavalisco, M., 
Conroy, C., Wechsler, R.~H., Ferguson, H.~C., Somerville, R.~S., Dickinson, 
M.~E., \& Urry, C.~M.\ 2009, \apj, 695, 368
\bibitem[Le F{\`e}vre et al.(2005)]{LeFevre05} Le F{\`e}vre, O., et al.\ 2005, \aap, 439, 877
\bibitem[Li et al.(2006)]{Li06} Li, C., Kauffmann, G., Jing, 
Y.~P., White, S.~D.~M., B{\"o}rner, G., \& Cheng, F.~Z.\ 2006, \mnras, 368, 21
\bibitem[Limber(1954)]{Limber54} Limber, D.~N.\ 1954, \apj, 119, 655
\bibitem[Loh et al.(2010)]{Loh10} Loh, Y.-S., et al.\ 2010, 
\mnras, 921
\bibitem[Ma \& Fry(2000)]{Ma00} Ma, C.-P., \& Fry, J.~N.\ 2000, \apj, 543, 503 
\bibitem[Madgwick et al.(2003)]{Madgwick03} Madgwick, D.~S., et al.\ 2003, \mnras, 344, 847
\bibitem[Mandelbaum et al.(2006)]{Mandelbaum06} Mandelbaum, R., 
Seljak, U., Kauffmann, G., Hirata, C.~M., \& Brinkmann, J.\ 2006, \mnras, 368, 715
\bibitem[Matsuoka et al.(2010)]{Matsuoka10} Matsuoka, Y., Masaki, S., Kawara, K., \& Sugiyama, N.\ 2010, arXiv:1008.0516
\bibitem[Marchesini et al.(2009)]{Marchesini09} Marchesini, D., van Dokkum, P.~G., F{\"o}rster Schreiber, N.~M., Franx, M., Labb{\'e}, I., \& Wuyts, S.\ 2009, \apj, 701, 1765
\bibitem[Maraston(2005)]{Maraston05} Maraston, C.\ 2005, \mnras, 362, 799
\bibitem[McCracken et al.(2008)]{McCracken08} McCracken, H.~J., Ilbert, O., Mellier, Y., Bertin, E., Guzzo, L., Arnouts, S., Le F{\`e}vre, O., \& Zamorani, G.\ 2008, \aap, 479, 321
\bibitem[McCracken et al.(2010)]{McCracken10} McCracken, H.~J., et al.\ 2010, \apj, 708, 202
\bibitem[Meneux et al.(2008)]{Meneux08} Meneux, B., et al.\ 2008, \aap, 478, 299
\bibitem[Meneux et al.(2009)]{Meneux09} Meneux, B., et al.\ 2009, \aap, 505, 463
\bibitem[More et al.(2010)]{More10} More, S., van den Bosch, 
F.~C., Cacciato, M., Skibba, R., Mo, H.~J., \& Yang, X.\ 2010, arXiv:1003.3203 
\bibitem[Moster et al.(2010)]{Moster10} Moster, B.~P., Somerville, R.~S., Maulbetsch, C., van den Bosch, F.~C., Macci{\`o}, A.~V., Naab, T., \& Oser, L.\ 2010, \apj, 710, 903
\bibitem[Muzzin et al.(2009)]{Muzzin09} Muzzin, A., Marchesini, 
D., van Dokkum, P.~G., Labb{\'e}, I., Kriek, M., \& Franx, M.\ 2009, \apj, 701, 1839
\bibitem[Myers et al.(2009)]{Myers09} Myers, A.~D., White, M., 
\& Ball, N.~M.\ 2009, \mnras, 399, 2279 
\bibitem[Navarro et al.(1996)]{Navarro96} Navarro, J.~F., Frenk, C.~S., \& White, S.~D.~M.\ 1996, \apj, 462, 563
\bibitem[Norberg et al.(2001)]{Norberg01} Norberg, P., et al.\ 
2001, \mnras, 328, 64
\bibitem[Norberg et al.(2002)]{Norberg02} Norberg, P., et al.\ 
2002, \mnras, 332, 827
\bibitem[Norberg et al.(2009)]{Norberg09} Norberg, P., Baugh, C.~M., Gazta{\~n}aga, E., \& Croton, D.~J.\ 2009, \mnras, 396, 19
\bibitem[Ouchi et al.(2005)]{Ouchi05} Ouchi, M., et al.\ 2005, \apjl, 635, L117
\bibitem[Peacock \& Smith(2000)]{Peacock00} Peacock, J.~A., \& Smith, R.~E.\ 2000, \mnras, 318, 1144
\bibitem[P{\'e}rez-Gonz{\'a}lez et al.(2008)]{Perez-Gonzalez08}P{\'e}rez-Gonz{\'a}lez, P.~G., et al.\ 2008, \apj, 675, 234 
\bibitem[Phleps et al.(2006)]{Phleps06} Phleps, S., Peacock, J.~A., Meisenheimer, K., \& Wolf, C.\ 2006, \aap, 457, 145
\bibitem[Pollo et al.(2006)]{Pollo06} Pollo, A., et al.\ 2006, \aap, 451, 409
\bibitem[Porciani \& Norberg(2006)]{Porciani06} Porciani, C., \& Norberg, P.\ 2006, \mnras, 371, 1824
\bibitem[Quadri et al.(2007)]{Quadri07} Quadri, R., et al.\ 
2007, \apj, 654, 138
\bibitem[Quadri et al.(2008)]{Quadri08} Quadri, R.~F., Williams, R.~J., Lee, K.-S., Franx, M., van Dokkum, P., \& Brammer, G.~B.\ 2008, \apjl, 685, L1 
\bibitem[Roche et al.(1999)]{Roche99} Roche, N., Eales, S.~A., Hippelein, H., \& Willott, C.~J.\ 1999, \mnras, 306, 538
\bibitem[Roche et al.(2002)]{Roche02} Roche, N.~D., Almaini, 
O., Dunlop, J., Ivison, R.~J., \& Willott, C.~J.\ 2002, \mnras, 337, 1282
\bibitem[Ross \& Brunner(2009)]{Ross09} Ross, A.~J., \& Brunner, R.~J.\ 2009, \mnras, 399, 878
\bibitem[Ross et al.(2010)]{Ross10} Ross, A.~J., Percival, W.~J., \& Brunner, R.~J.\ 2010, \mnras, 879
\bibitem[Sanders et al.(2007)]{Sanders07} Sanders, D.~B., et al.\ 2007, \apjs, 172, 86
\bibitem[Seljak(2000)]{Seljak00} Seljak, U.\ 2000, \mnras, 318, 203
\bibitem[Seljak \& Warren(2004)]{Seljak04} Seljak, U., \& Warren, M.~S.\ 2004, \mnras, 355, 129
\bibitem[Simon et al.(2009)]{Simon09} Simon, P., Hetterscheidt, 
M., Wolf, C., Meisenheimer, K., Hildebrandt, H., Schneider, P., Schirmer, 
M., \& Erben, T.\ 2009, \mnras, 398, 807
\bibitem[Scoccimarro et al.(2001)]{Scoccimarro01} Scoccimarro, R., Sheth, R.~K., Hui, L., \& Jain, B.\ 2001, \apj, 546, 20
\bibitem[Scoville et al.(2007)]{Scoville07} Scoville, N., et al.\ 2007, \apjs, 172, 1
\bibitem[Sheth \& Tormen(1999)]{Sheth99} Sheth, R.~K., \& Tormen, G.\ 1999, \mnras, 308, 119
\bibitem[Sheth et al.(2001)]{Sheth01} Sheth, R.~K., Mo, H.~J., \& Tormen, G.\ 2001, \mnras, 323, 1
\bibitem[Skibba, Sheth \& Martino(2007)]{Skibba07} Skibba, R. A., Sheth, R.~K., \& Martino, M. C.\ 2007, MNRAS, 382, 1940
\bibitem[Springel et al.(2005)]{Springel05} Springel, V., et al.\ 2005, \nat, 435, 629
\bibitem[Stott et al.(2010)]{Stott10} Stott, J.~P., et al.\ 2010, \apj, 718, 23
\bibitem[Swanson et al.(2008)]{Swanson08} Swanson, M.~E.~C., 
Tegmark, M., Blanton, M., \& Zehavi, I.\ 2008, \mnras, 385, 1635
\bibitem[Tinker et al.(2005)]{Tinker05} Tinker, J.~L., Weinberg, D.~H., Zheng, Z., \& Zehavi, I.\ 2005, \apj, 631, 41
\bibitem[Tinker et al.(2010a)]{Tinker10a} Tinker, J.~L., Wechsler, R.~H., \& Zheng, Z.\ 2010, \apj, 709, 67
\bibitem[Tinker et al.(2010b)]{Tinker10b} Tinker, J.~L., Robertson, B.~E., Kravtsov, A.~V., Klypin, A., Warren, M.~S., Yepes, G., \& Gottlober, S.\ 2010, arXiv:1001.3162
\bibitem[Vale \& Ostriker(2006)]{Vale06} Vale, A., \& Ostriker, J.~P.\ 2006, \mnras, 371, 1173 
\bibitem[van der Wel et al.(2006)]{Wel06} van der Wel, A., Franx, M., Wuyts, S., van Dokkum, P.~G., Huang, J., Rix, H.-W., \& Illingworth, G.~D.\ 2006, \apj, 652, 97
\bibitem[van Dokkum et al.(2006)]{Dokkum06} van Dokkum, P.~G., et al.\ 2006, \apjl, 638, L59
\bibitem[van Dokkum et al.(2009)]{Dokkum09} van Dokkum, P.~G., et al.\ 2009, \pasp, 121, 2
\bibitem[Wake et al.(2008a)]{Wake08a} Wake, D.~A., et al.\ 2008, 
\mnras, 387, 1045
\bibitem[Wake et al.(2008b)]{Wake08b} Wake, D.~A., Croom, S.~M., 
Sadler, E.~M., \& Johnston, H.~M.\ 2008, \mnras, 391, 1674
\bibitem[Wetzel et al.(2007)]{Wetzel07} Wetzel, A.~R., Cohn, J.~D., White, M., Holz, D.~E., \& Warren, M.~S.\ 2007, \apj, 656, 139
\bibitem[Whitaker et al.(2010a)]{Whitaker10} Whitaker, K.~E., et 
al.\ 2010, \apj, 719, 1715
\bibitem[Whitaker et al.(2010b)]{Whitaker10b} Whitaker, K.~E., et 
al.\ 2010, \apj, submitted
\bibitem[White et al.(2007)]{White07} White, M., Zheng, Z., 
Brown, M.~J.~I., Dey, A., \& Jannuzi, B.~T.\ 2007, \apjl, 655, L69
\bibitem[White \& Rees(1978)]{White78} White, S.~D.~M., \& Rees, M.~J.\ 1978, \mnras, 183, 341
\bibitem[Yan et al.(2003)]{Yan03} Yan, R., Madgwick, D.~S., 
\& White, M.\ 2003, \apj, 598, 848
\bibitem[Yang et al.(2003)]{Yang03} Yang, X., Mo, H.~J., \& van den Bosch, F.~C.\ 2003, \mnras, 339, 1057
\bibitem[Yang et al.(2005)]{Yang05} Yang, X., Mo, H.~J., Jing, 
Y.~P., \& van den Bosch, F.~C.\ 2005, \mnras, 358, 217
\bibitem[Yang et al.(2008)]{Yang08} Yang, X., Mo, H.~J., \& van den Bosch, F.~C.\ 2008, \apj, 676, 248 
\bibitem[York et al.(2000)]{York00} York, D.~G., et al.\ 2000, \aj, 120, 1579
\bibitem[Zehavi et al.(2002)]{Zehavi02} Zehavi, I., et al.\ 
2002, \apj, 571, 172
\bibitem[Zehavi et al.(2004)]{Zehavi04} Zehavi, I., et al.\ 
2004, \apj, 608, 16
\bibitem[Zehavi et al.(2005)]{Zehavi05} Zehavi, I., et al.\ 2005, \apj, 630, 1
\bibitem[Zehavi et al.(2010)]{Zehavi10} Zehavi, I., et al.\ 2010,  arXiv:1005.2413
\bibitem[Zheng et al.(2005)]{Zheng05} Zheng, Z., et al.\ 2005, \apj, 633, 791
\bibitem[Zheng et al.(2007)]{Zheng07} Zheng, Z., Coil, A.~L., \& Zehavi, I.\ 2007, \apj, 667, 760
\end{thebibliography}
\end{document}